\documentclass[aps,prb,superscriptaddress,floatfix,amsmath,amsfonts,twocolumn]{revtex4-2}  
\usepackage[russian,english]{babel}

\usepackage{enumitem}
\usepackage{amsmath}\usepackage{graphicx}
\usepackage{epsfig}
\usepackage{bm}
\usepackage{bbold}
\usepackage{amssymb}
\usepackage{upgreek}
\usepackage{color}
\usepackage[normalem]{ulem}
\usepackage[colorlinks=true,citecolor=blue,linkcolor=blue,urlcolor=blue]{hyperref}

\begin{document}
\title{Acoustic spin resonance in polariton condensates}

\author{D. A. Saltykova}
\affiliation{Department of Physics, ITMO University, Saint Petersburg 197101, Russia}

\author{A. Kudlis}
\affiliation{Science Institute, University of Iceland, Dunhagi 3, IS-107, Reykjavik, Iceland}

\author{A. V. Yulin}
\affiliation{Department of Physics, ITMO University, Saint Petersburg 197101, Russia}

\author{I. A. Shelykh}
\affiliation{Science Institute, University of Iceland, Dunhagi 3, IS-107, Reykjavik, Iceland}

\begin{abstract}
We theoretically investigate acoustic spin resonance in a spatially homogeneous
spinor polariton condensate. A longitudinal acoustic wave generates a
time-periodic strain-induced effective magnetic field acting on the condensate
pseudospin. When this field is transverse to the static in-plane
linear-polarization splitting, it resonantly drives polarization oscillations.
We show that spin-dependent interactions shift the resonance and produce
nonlinear line shapes, while gain, reservoir dynamics, and spin relaxation make
the response dissipative and history-dependent, producing amplitude hysteresis. In the presence of lifetime anisotropy, the condensate can develop a
bifurcated stationary state with finite circular polarization, and a resonant
acoustic drive can switch between the corresponding out-of-plane branches. A
Zeeman splitting provides an additional conservative knob for tuning the
resonance frequency. Our results identify coherent acoustic driving as a route
to resonant, nonlinear, and switchable control of polariton pseudospin dynamics.
\end{abstract}

\maketitle

\section{Introduction}

Exciton--polaritons are hybrid light--matter quasiparticles emerging in the
strong-coupling regime between a photonic mode of a planar semiconductor
microcavity and an excitonic resonance in a quantum well embedded in the
antinode of the cavity mode. From their photonic component, polaritons inherit
an extremely small effective mass, about \(10^{-5}\) of the free-electron mass,
and a large coherence length on the micrometer scale \cite{Ballarini2017}. The
excitonic component makes polariton systems sensitive to external electric and
magnetic fields and gives rise to robust polariton--polariton interactions
\cite{Glazov2009}. This combination of properties allows one to observe
collective quantum phenomena such as polariton Bose--Einstein condensation,
superfluidity, and polariton lasing in semiconductor microcavities; see Refs.~\cite{Deng2010,Carusotto2013} for reviews.

An important characteristic of cavity polaritons is their spin, or pseudospin,
structure \cite{ShelykhReview}. As for photons, polaritons have two spin
projections on the structure growth axis, corresponding to the two opposite
circular polarizations. Polariton--polariton interactions, provided mainly by
the exchange interaction of their excitonic components, are spin-dependent
\cite{Ciuti1998}: polaritons with the same circular polarization interact much
more strongly than polaritons with opposite circular polarizations
\cite{Glazov2009}. This spin anisotropy contributes to the spontaneous build-up
of linear polarization in polariton Bose--Einstein condensates
\cite{Baumberg2008} and to self-induced Larmor precession of polarization in
elliptically pumped polariton condensates \cite{ShelykhSILP}.

Static splittings of the spinor polariton doublet provide effective magnetic
fields acting on the pseudospin. A magnetic field applied along the growth axis
splits the two circular polarizations and produces a Zeeman term. In contrast,
anisotropies of the cavity structure produce splittings in the
linear-polarization basis and therefore act as in-plane effective fields. These
include the \(k\)-dependent TE--TM splitting of the photonic modes of a planar
resonator, which plays the role of an effective spin--orbit interaction
\cite{ShelykhReview}, and \(k\)-independent contributions caused by intrinsic
birefringence, residual strain, or other structural anisotropies of the cavity
material \cite{Klopotowski2006,Balili2010}. The latter define a constant in-plane effective magnetic field whose
orientation is determined by the principal optical axes of the microcavity.
Depending on the balance between precession, gain, losses, and spin relaxation,
such fields can lead either to pseudospin precession or to polarization pinning,
as observed in polariton condensates \cite{Kasprzak2006,Kasprzak2007}.

Acoustic waves provide a complementary route to polariton control. Surface
and bulk acoustic waves can modulate polariton energies and couplings through
deformation-potential, piezoelectric, photoelastic, and optomechanical
mechanisms, enabling acoustic Bragg scattering, dynamic polariton lattices,
and coherent polariton--phonon coupling
\cite{Ivanov2003,cho2005bragg,lanzillotti2007coherent,CerdaMendez2010,vishnevsky2011coherent,Kuznetsov2019,Chafatinos2020,RamosPerez2024}.
 It was also demonstrated that acousto-polariton interactions can lead to
parametric generation of new frequencies in semiconductor microcavities \cite{yulin2019resonant,carraro2024solid}. 

In many of these settings, the acoustic wave acts primarily as a scalar
modulation of the polariton energy or of the coupling between modes. Here we focus instead on
the polarization-dependent strain response, where the acoustic wave acts as a
time-periodic in-plane effective magnetic field for the polariton pseudospin.

In this work we develop a minimal theory of acoustic spin resonance in a
spatially homogeneous polariton condensate. We show how interactions, reservoir
dynamics, lifetime anisotropy, and Zeeman splitting transform the basic
transverse-drive resonance into nonlinear, hysteretic, switchable, and tunable
pseudospin dynamics.

\section{Model}

We consider a spatially homogeneous spinor polariton condensate described by
the macroscopic wavefunction \(\psi_\sigma(t)\), \(\sigma=\pm1\), coupled to an
incoherent excitonic reservoir with density \(n_R(t)\). For simplicity, the
reservoir is assumed to be spin-unpolarized. The condensate pseudospin is
introduced as

\begin{equation}
S_x=2\,{\rm Re}(\psi_+^\ast\psi_-), \
S_y=2\,{\rm Im}(\psi_+^\ast\psi_-), \
S_z=|\psi_+|^2-|\psi_-|^2 .
\label{eq:pseudospin_definition}
\end{equation}
The pseudospin determines the condensate polarization state: the \(Z\) axis
corresponds to circular polarizations, while the \(X\) and \(Y\) axes correspond
to two pairs of orthogonal linear polarizations. Its length is the total
condensate density,
\[
|\mathbf S|=|\psi_+|^2+|\psi_-|^2\equiv n .
\]

Throughout Eqs.~\eqref{Sdyn}--\eqref{nRdyn}, the spinor amplitudes,
the pseudospin \(\mathbf S\), and the density \(n=|\mathbf S|\) are
 dimensionless. Physical areal densities are denoted by capital letters. The coupled condensate--reservoir dynamics is written in the pseudospin form as
\begin{widetext}
\begin{align}
\partial_t \mathbf{S}
&=
(Rn_R-\gamma)\mathbf{S}
-\gamma_a |\mathbf{S}|\,\mathbf e_x
+
\left[\mathbf{B}_0(\mathbf{S})+\mathbf{B}_a(t)\right]\times \mathbf{S}
-\lambda\, \mathbf{S} \times
\left[\mathbf{B}_0(\mathbf{S}) \times \mathbf{S} \right],
\label{Sdyn}
\\
\partial_t n_R
&=
P-\gamma_R n_R-Rn_R|\mathbf{S}| .
\label{nRdyn}
\end{align}
\end{widetext}
The derivation of Eqs.~\eqref{Sdyn}--\eqref{nRdyn} is given in
Supplementary Note~1 of the Supplemental Material. Here \(R\) is the stimulated scattering rate from the reservoir to the
condensate, \(\gamma\) is the polarization-independent condensate decay rate,
\(P\) is the incoherent pump, and \(\gamma_R\) is the reservoir decay rate. The
parameter \(\gamma_a\) describes lifetime anisotropy of two orthogonal linearly
polarized modes. With our sign convention, the corresponding term in
Eq.~\eqref{Sdyn} favors states with negative \(S_x\), in analogy with the
phenomenological description used in Ref.~\cite{Ohadi2015}.

\begin{figure}[b]
    \centering
    \includegraphics[width=0.9\columnwidth]{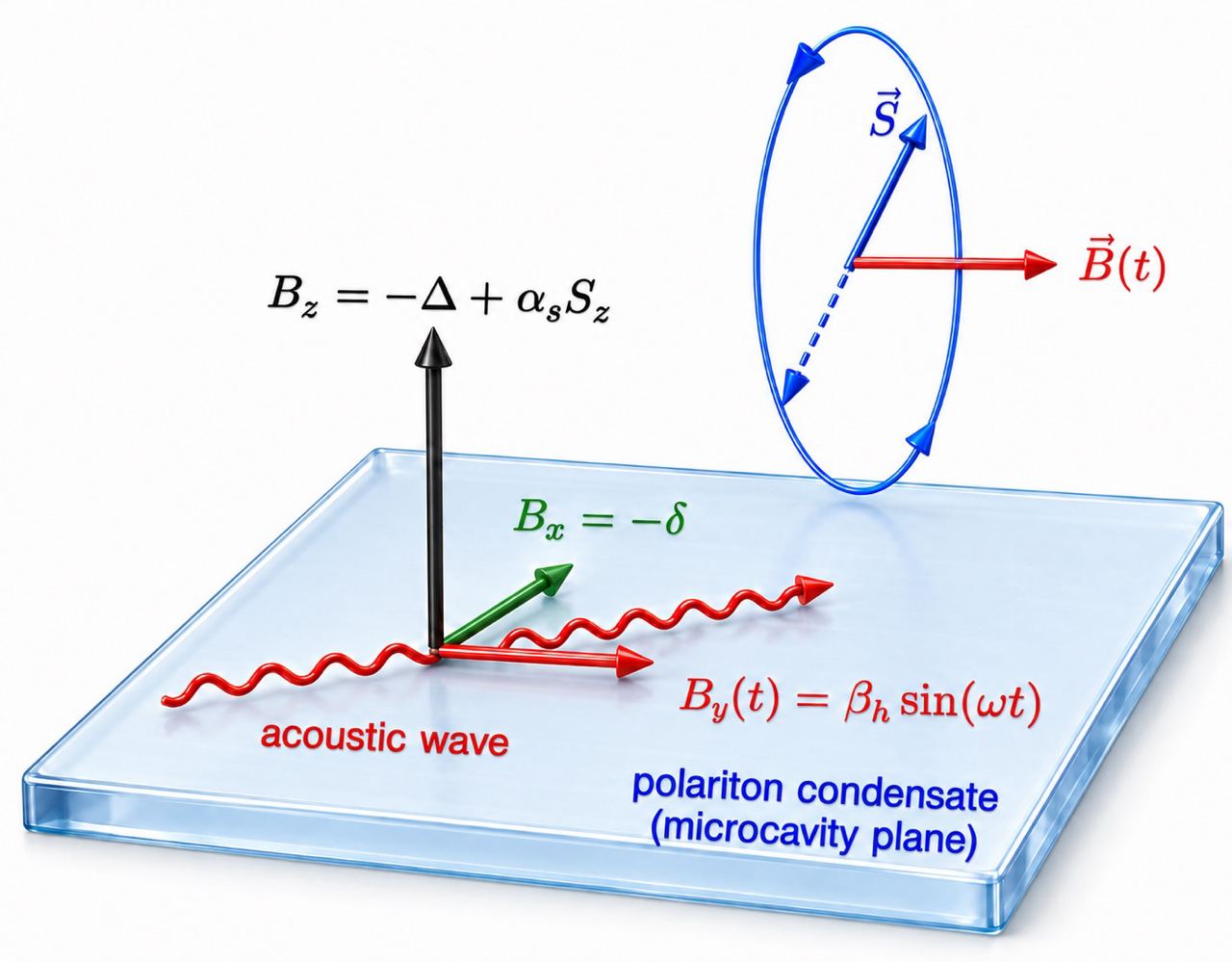}
    \caption{\textbf{Geometry of acoustic spin resonance in a polariton
    condensate.}
    A longitudinal acoustic wave propagating in the microcavity plane produces a
    time-periodic strain-induced effective magnetic field acting on the
    condensate pseudospin \(\mathbf S\). The static field contains the
    in-plane splitting \(B_x=-\delta\) and the out-of-plane component
    \(B_z=-\Delta+\alpha_sS_z\), which combines the Zeeman term and the
    self-induced field. In the resonant geometry used in the
    main simulations, \(\phi=\pi/4\), the acoustic contribution is transverse to
    the static in-plane field and is written as
    \(B_y(t)=\beta_h\sin(\omega t)\). The acoustic drive can therefore excite
    pseudospin precession on the Poincar\'e sphere and generate large
    polarization oscillations.}
    \label{fig:setup}
\end{figure}

The stationary part of the effective magnetic field is
\begin{equation}
\mathbf{B}_0(\mathbf{S})
=
-\delta\,\mathbf e_x
+
\left(\alpha_s S_z-\Delta\right)\mathbf e_z .
\label{eq:static_field}
\end{equation}
Here \(\delta\) is the static linear-polarization splitting caused by intrinsic
birefringence, \(\Delta\) is the circular-polarization splitting induced by an
external magnetic field, and \(\alpha_s=\alpha_1-\alpha_2\) is the difference
between the interaction constants for polaritons with the same and opposite
circular polarizations. The term \(\alpha_sS_z\mathbf e_z\) is the
self-induced field produced by spin-dependent interactions.

The acoustic wave induces the time-dependent field
\begin{equation}
\mathbf{B}_a(t)
=
\beta_h f_{\rm on}(t)
\left(\mathbf e_x\cos 2\phi+\mathbf e_y\sin 2\phi\right)
\sin\Theta(t).
\label{eq:acoustic_field}
\end{equation}
The parameter \(\beta_h\) is proportional to the acoustic amplitude, \(\phi\) is
the propagation angle of the acoustic wave in the microcavity plane, and
\(\Theta(t)\) is the acoustic phase. For a monochromatic drive,
\(\Theta(t)=\omega t\); for a slow frequency sweep,
\(\dot\Theta(t)=\omega(t)\). In fixed-frequency scans the drive is switched on
smoothly by
\begin{equation}
f_{\rm on}(t)
=
\frac12
\left[
1+\tanh\left(\frac{t-5\tau_{\rm on}}{\tau_{\rm on}}\right)
\right].
\label{eq:turn_on}
\end{equation}
The single-mode approximation used below neglects the spatial variation of the
acoustic strain across the condensate. This approximation is valid when the
condensate size is smaller than the acoustic wavelength, or more generally when
the driven mode samples only a weakly varying part of the strain profile.

The last term in Eq.~\eqref{Sdyn} is a Gilbert-type spin-relaxation term. It is
the isotropic part of the spin-relaxation mechanism discussed in
Ref.~\cite{Saltykova2025spin}. This term damps pseudospin motion transverse to
the local spin eigenaxis and drives the condensate toward the locally stable
spin eigenaxis associated with \(\mathbf B_0(\mathbf S)\), while conserving the
condensate density by itself. In the present minimal model we retain this
isotropic contribution and write it with respect to the static field
\(\mathbf B_0(\mathbf S)\). The additional anisotropic spin-relaxation force, as
well as corrections obtained by including the weak acoustic field in the
relaxation channel, are neglected since they change relaxation rates but do not
qualitatively affect the acoustic-resonance mechanisms discussed below. In the
figures this parameter is denoted by
\(\lambda_{\rm relax}\equiv\lambda\).

The driven response is quantified by the oscillation amplitude of a normalized
pseudospin component,
\begin{equation}
A_j(W)
=
\frac12
\left[
\max_{t\in W}\frac{S_j(t)}{|\mathbf S(t)|}
-
\min_{t\in W}\frac{S_j(t)}{|\mathbf S(t)|}
\right],
\qquad j=x,y,z .
\label{eq:response_amplitude}
\end{equation}
For fixed-frequency scans, the measurement window \(W\) is taken after the
turn-on transient. For slow sweeps, \(W\) is a sliding time window. We use \(A_x\)
to characterize the response of the linear-polarization component and \(A_z\)
to characterize the circular-polarization response. 

For dimensional plots we use the conversion convention
\begin{equation}
\begin{gathered}
E_0\omega=\hbar\Omega,\qquad
\delta_E=E_0\delta,\qquad
\Delta_Z=E_0\Delta,\\
\beta_{h,E}=E_0\beta_h,\qquad
\gamma_{a,E}=E_0\gamma_a .
\end{gathered}
\label{eq:unit_conversion}
\end{equation}
Here \(E_0\simeq0.0207~\mathrm{meV}\), \(\Omega\) is the physical acoustic
angular frequency, while \(\omega\) is the dimensionless drive frequency used in the simulations. Thus, when an axis is labelled by \(\omega\) in
\(\upmu\mathrm{eV}\), it displays the corresponding acoustic drive energy. The derivation of this conversion and the definition of the dimensionless units
are given in Supplementary Note~2 of the Supplemental Material.

\section{Results and discussion}

\subsection{Conservative acoustic spin resonance}

\begin{figure}[t]
    \centering
    \includegraphics[width=\columnwidth]{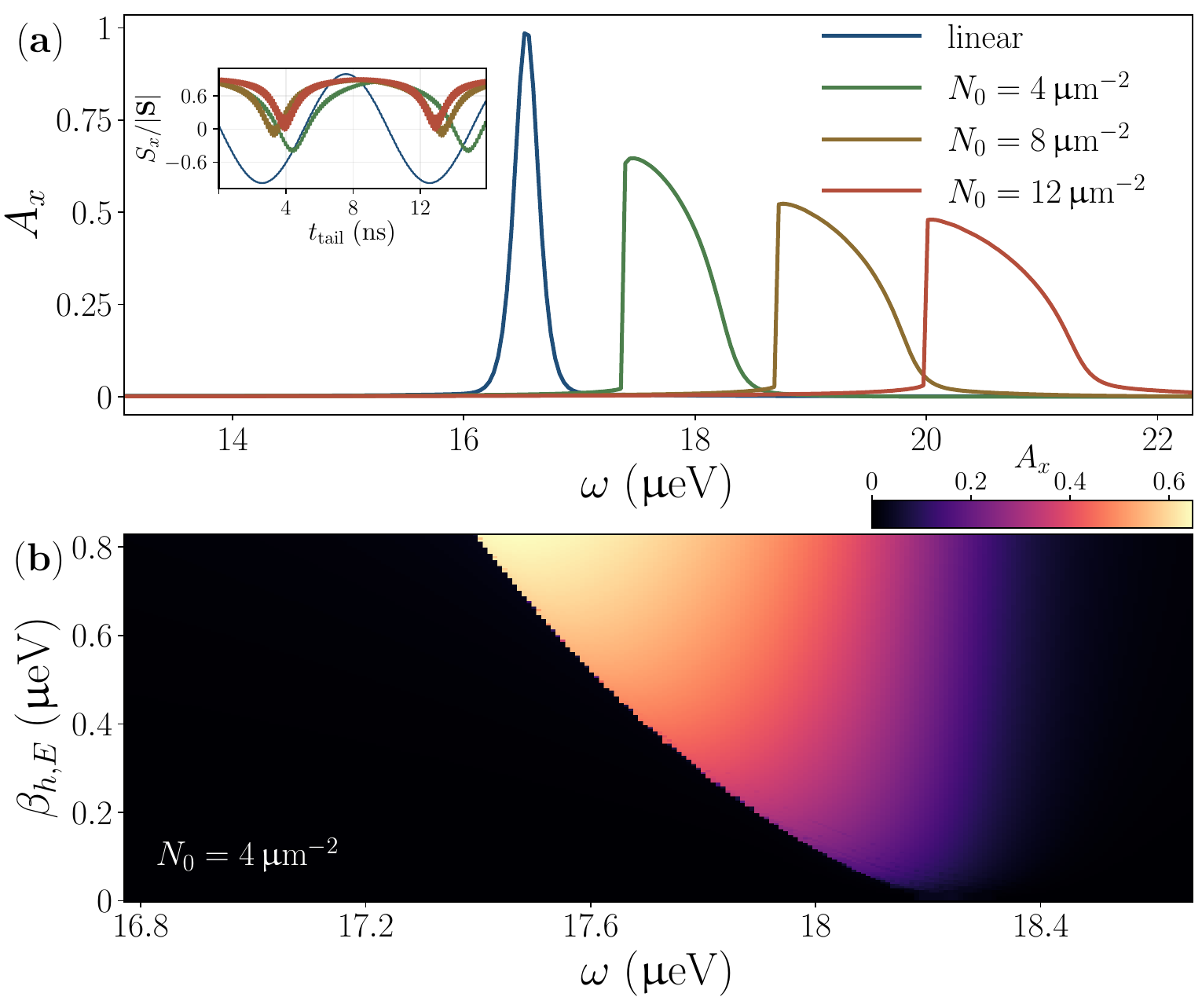}
    \caption{{\textbf{Conservative acoustic pseudospin resonance.}
    Calculated from the conservative limit of Eq.~\eqref{Sdyn}, with gain,
    losses, reservoir coupling, lifetime anisotropy, Zeeman splitting, and spin
    relaxation neglected. The acoustic wave propagates at \(\phi=\pi/4\), so
    that the acoustic field is transverse to the static in-plane splitting. We
    use \(E_0=0.020678~\mathrm{meV}\), \(E_0/h=5.00~\mathrm{GHz}\),
    \(t_0=31.83~\mathrm{ps}\), and
    \(\delta_E=16.54~\upmu\mathrm{eV}\). The nonlinear curves are calculated
    with \(\alpha_s=0.35\), corresponding to
    \(\alpha_{\rm phys}=0.905~\upmu\mathrm{eV}\,\upmu\mathrm{m}^2\) and
    \(n_*^{(c)}=8.0~\upmu\mathrm{m}^{-2}\). The labels
    \(N_0=4,8,12~\upmu\mathrm{m}^{-2}\) denote physical condensate densities;
    the corresponding conserved dimensionless spin lengths are
    \(N_0/n_*^{(c)}=0.5,1.0,1.5\).}
    (a) Frequency response \(A_x\) versus the acoustic drive energy
    \(E_0\omega=\hbar\Omega\) at
    \(\beta_{h,E}=0.827~\upmu\mathrm{eV}\). The blue curve is the
    linear reference \((\alpha_s=0)\), while the nonlinear curves correspond to
    \(N_0=4,8,12~\upmu\mathrm{m}^{-2}\). The inset shows representative
    late-time traces of \(S_x/|\mathbf S|\).
    (b) Response map \(A_x(\omega,\beta_{h,E})\), with \(\omega\) displayed
    through \(E_0\omega\), for \(N_0=4~\upmu\mathrm{m}^{-2}\). The
    strong-response region is shifted and broadened by the nonlinear
    self-induced field.}
    \label{fig:conservative_case}
\end{figure}

We first consider the conservative limit of Eq.~\eqref{Sdyn}. Gain, radiative
losses, reservoir coupling, lifetime anisotropy, Zeeman splitting, and spin
relaxation are neglected, so the reservoir equation drops out and
the spin length
\(n_0^{(c)}\equiv |\mathbf S|\) is conserved. The corresponding physical
condensate density is
\(N_0=n_*^{(c)}n_0^{(c)}\). Thus \(N_0\) in
Fig.~\ref{fig:conservative_case} denotes a physical areal density, while the
conserved dimensionless spin lengths are \(n_0^{(c)}=0.5,1.0,1.5\) for
\(N_0=4,8,12~\upmu\mathrm{m}^{-2}\), respectively.

The acoustic wave is chosen to propagate at \(45^\circ\) with respect to the
\(X\) axis. According to Eq.~\eqref{eq:acoustic_field}, this makes the acoustic
field directed along \(Y\), transverse to the static linear-polarization
splitting along \(X\). The same conservative dynamics written in terms of the circularly polarized
components \(\psi_\pm\) is given in Supplementary Note~3 of the Supplemental
Material. In the linear limit, the problem reduces to the standard
magnetic-resonance geometry: a pseudospin in a stationary field is driven by a
weak transverse oscillating field. Interactions modify this picture because the
field \(\alpha_s S_z\mathbf e_z\) depends on the instantaneous pseudospin state.

Figure~\ref{fig:conservative_case}(a) shows the finite-time frequency response
\(A_x\) as a function of the acoustic drive energy \(E_0\omega=\hbar\Omega\),
where \(\omega\) is the dimensionless drive frequency used in the simulations.
The linear reference, obtained by setting \(\alpha_s=0\), has a sharp
finite-time resonance peak close to the static splitting. For nonzero density,
the self-induced field shifts the strongest response to higher energies and
makes the line shape asymmetric. The inset illustrates the corresponding
late-time traces of \(S_x/|\mathbf S|\): in the nonlinear regime the response is
no longer purely harmonic but develops a slow modulation due to feedback between
the drive and the interaction-induced field. Figure~\ref{fig:conservative_case}(b)
shows that the strong-response region occupies a finite domain in the
\((\omega,\beta_{h,E})\) plane, with \(\omega\) displayed through
\(E_0\omega\) and \(\beta_{h,E}=E_0\beta_h\). Since the system is conservative,
this onset should not be interpreted as a dissipative threshold; it reflects the
nonlinear efficiency of acoustic coupling to the interaction-renormalized
pseudospin motion.

\subsection{Minimal pumped--dissipative response and amplitude hysteresis}

We now restore gain, losses, reservoir coupling, and spin relaxation, but keep
\(\Delta=0\) and \(\gamma_a=0\). Thus no external magnetic field and no
polarization-dependent lifetime anisotropy are present in this subsection. The
condensation threshold of the isotropic-loss model is
\[
P_{\rm th}=\frac{\gamma\gamma_R}{R}.
\]
Above threshold, the reservoir fixes the characteristic condensate density,
while gain, losses, and spin relaxation provide dissipative channels through
which the pseudospin can approach long-time driven states. Hence, in contrast
to the conservative case, the driven system can relax to attractors selected by
the acoustic forcing. This makes it possible to observe not only a resonance
shift, but also multistability and history-dependent response.

\begin{figure}[t]
    \centering
    \includegraphics[width=\columnwidth]{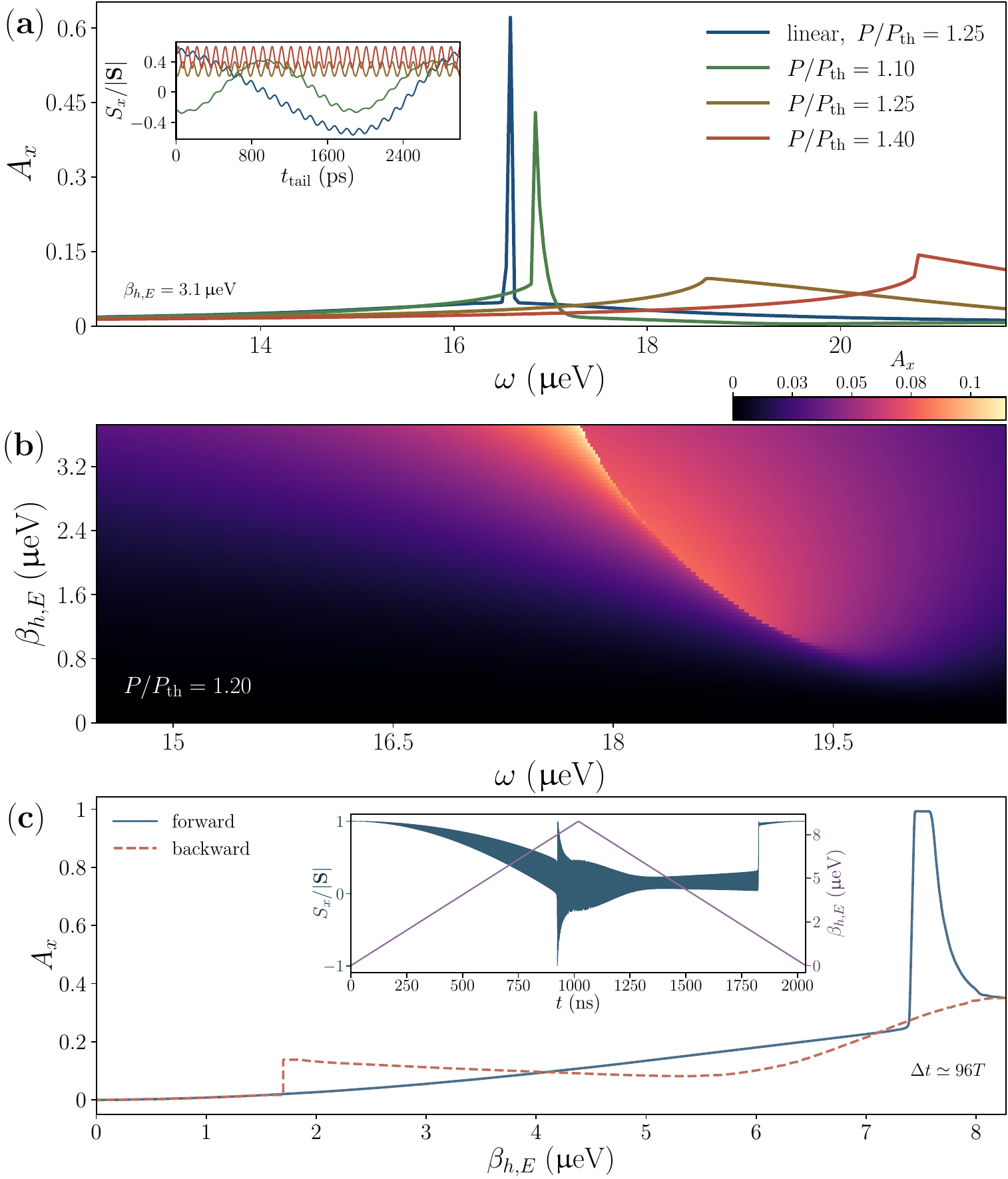}
    \caption{\textbf{Dissipative acoustic response and slow-amplitude
hysteresis.}
All panels are calculated from Eqs.~\eqref{Sdyn}--\eqref{nRdyn} for the
minimal pumped--dissipative specialization with
\(\gamma_a=\Delta=0\), \(\lambda_{\rm relax}=0.05\), and
\(\phi=\pi/4\). The nonlinear curves, maps, and sweeps use
\(\alpha_s=0.5\); the blue curve in panel (a) is the linear reference
\((\alpha_s=0)\). The dimensional calibration is
\(E_0=0.020678~\mathrm{meV}\), so that
\(E_0/h=5.00~\mathrm{GHz}\).
The driven simulations are initialized from the undriven linearly
polarized pumped state
\((S_x,S_y,S_z,n_R)(0)=(P-P_{\rm th},0,0,1)\), in dimensionless units.
(a) Frequency response \(A_x\) at fixed acoustic amplitude
\(\beta_{h,E}=3.10~\upmu\mathrm{eV}\). The blue curve is the
linear reference at \(P/P_{\rm th}=1.25\), while the nonlinear curves
correspond to \(P/P_{\rm th}=1.10,1.25,1.40\). The inset shows
representative late-time traces of \(S_x/|\mathbf S|\).
(b) Response map \(A_x(\omega,\beta_{h,E})\), with \(\omega\) displayed
through \(E_0\omega\), at \(P/P_{\rm th}=1.20\), showing the
amplitude-dependent nonlinear resonance region.
(c) Slow triangular sweep of the acoustic amplitude at
\(P/P_{\rm th}=1.25\) and fixed drive energy
\(E_0\omega=17.8~\upmu\mathrm{eV}\):
\(\beta_{h,E}:0\rightarrow8.27~\upmu\mathrm{eV}\rightarrow0\).
For the sweep this gives the initial state
\((S_x,S_y,S_z,n_R)(0)=(1,0,0,1)\); no reinitialization is performed between
the forward and backward parts of the triangular cycle.
The forward and backward branches are extracted from the time trace using a
sliding max--min window of approximately \(96\) acoustic periods. The separation
of the two branches demonstrates path-dependent switching between weak- and
strong-response attractors; the inset shows the corresponding
\(S_x/|\mathbf S|\) trace and the imposed amplitude protocol.}
    \label{fig:dissipative_overview}
\end{figure}

The frequency response in this minimal pumped--dissipative regime is shown in
Fig.~\ref{fig:dissipative_overview}(a). The blue curve is the linear reference
obtained for \(\alpha_s=0\). It peaks close to the static linear-polarization
splitting, as expected for a weakly driven pseudospin precessing in a fixed
in-plane field. For \(\alpha_s\neq0\), increasing the pump increases the
condensate density and therefore the accessible scale of the nonlinear field
\(\alpha_s S_z\). The corresponding small-amplitude estimate,
\(\omega_s^2=\delta(\delta+\alpha_s n_0)\), is derived in Supplementary
Note~4 of the Supplemental Material. The resonance is shifted to higher acoustic energies because
the interaction-induced out-of-plane field increases the effective precession
frequency once finite circular-polarization oscillations are generated. The
peak height is not a monotonic function of pump power, since \(A_x\) measures a
finite-amplitude nonlinear trajectory rather than a linear susceptibility. In
particular, stronger interactions can both enhance the coupling to the driven
motion and detune the pseudospin away from the optimal resonant condition.

Figure~\ref{fig:dissipative_overview}(b) shows the response map at fixed pump.
The strong-response domain bends in the \((\omega,\beta_{h,E})\) plane, with
\(\omega\) displayed through \(E_0\omega\) and
\(\beta_{h,E}=E_0\beta_h\). This bending is a clear manifestation of an amplitude-dependent resonance
condition. At weak acoustic amplitudes the system
responds close to the small-oscillation frequency around the linearly polarized
state. At larger amplitudes, the pseudospin explores a wider region of the
Poincar\'e sphere, so the nonlinear field \(\alpha_s S_z\mathbf e_z\) changes
during the motion and renormalizes the instantaneous precession frequency. As a
result, the acoustic amplitude itself becomes a control parameter for the
resonance.

Panel~\ref{fig:dissipative_overview}(c) demonstrates the corresponding memory
effect directly. During a slow triangular sweep of acoustic amplitude, the
upward and downward branches of \(A_x\) do not coincide. On the upward sweep, the system first follows a weak-response attractor and then
jumps to a large-amplitude driven state once the trajectory enters the nonlinear
transition region. On the downward sweep, the large-amplitude state persists down to
smaller acoustic amplitudes before the system returns to the weak-response
branch. Thus the same value of \(\beta_{h,E}\) can correspond to two different
long-time pseudospin motions, depending on the previous history of the drive.
The loop in Fig.~\ref{fig:dissipative_overview}(c) therefore represents
amplitude hysteresis of the acoustic polarization response rather than a
threshold in the pump.

\subsection{Lifetime-anisotropy-induced bifurcation and acoustic switching}

\begin{figure}[t]
    \centering
    \includegraphics[width=\columnwidth]{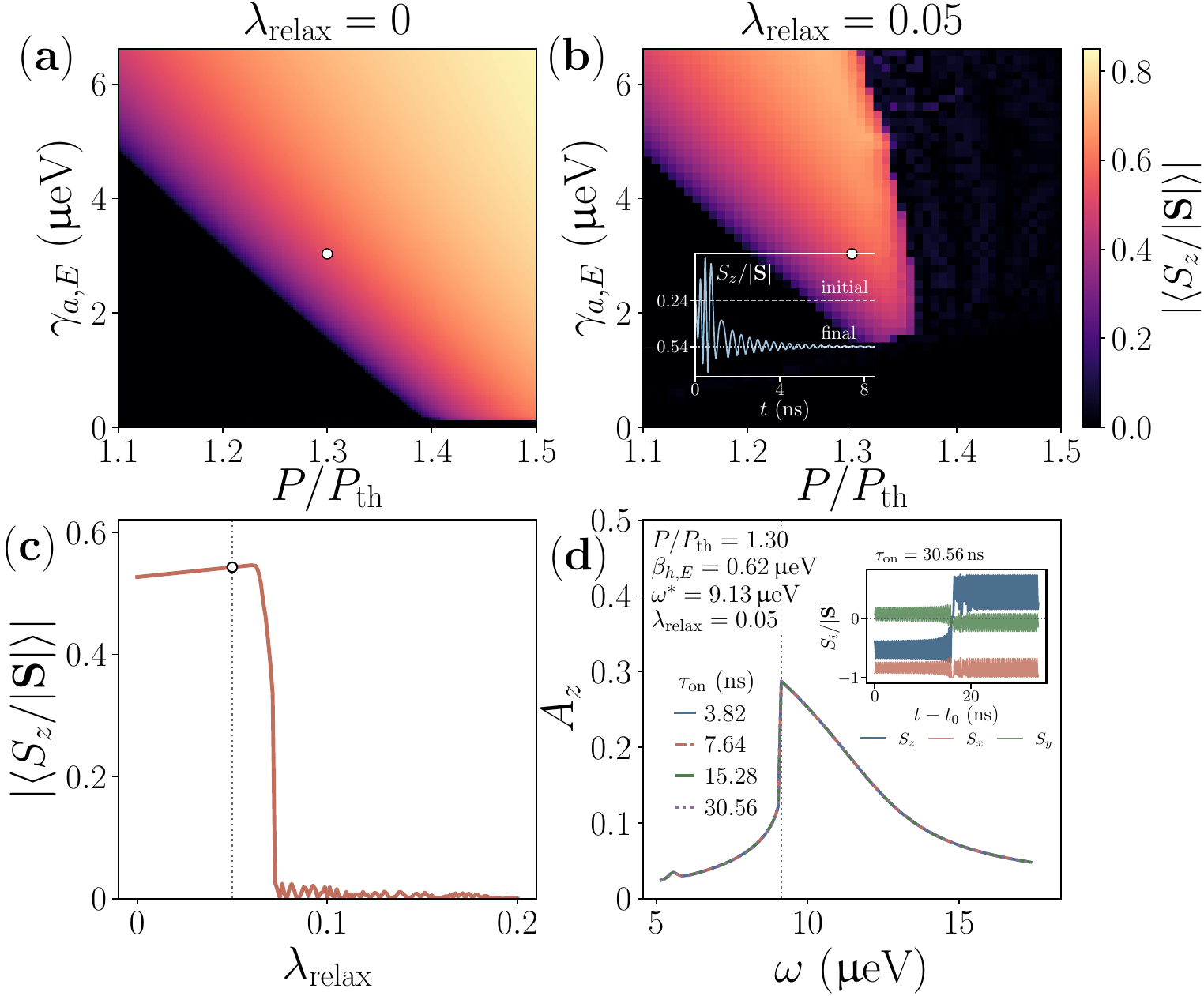}
    \caption{\textbf{Bifurcated stationary state and resonant acoustic response
induced by lifetime anisotropy.}
The calculations use \(\Delta=0\), \(\alpha_s=0.5\),
\(\delta_E=16.54~\upmu\mathrm{eV}\), and
\(E_0=0.020678~\mathrm{meV}\).
(a,b) Stationary maps of
\(|\langle S_z/|\mathbf S|\rangle|\) in the
\((P/P_{\rm th},\gamma_{a,E})\) plane without acoustic drive, for
\(\lambda_{\rm relax}=0\) and \(\lambda_{\rm relax}=0.05\), respectively.
The white markers indicate the same working point,
\(P/P_{\rm th}=1.30\) and
\(\gamma_{a,E}=3.03~\upmu\mathrm{eV}\).
The inset in panel (b) shows the relaxation toward the stationary out-of-plane
state at this working point.
The branch used for the driven calculations is selected by an undriven
preparatory relaxation at \(\beta_h=0\), starting from
$(S_x,S_y,S_z,n_R)(0)\simeq(0.12,0,0.03,1.30), \
\frac{\mathbf S(0)}{|\mathbf S(0)|}\simeq(0.970,0,0.243).$
Thus the seed is mostly linearly polarized and is not an ideal
\(z\)-polarized state. The relaxation reaches the stationary branch
$(S_x,S_y,S_z,n_R)_{\rm st}\simeq(-1.610,0.162,-1.046,0.877), \ \frac{\mathbf S_{\rm st}}{|\mathbf S_{\rm st}|}
\simeq(-0.836,0.084,-0.543).$
The opposite branch is obtained by reversing the sign of the initial
circular-polarization seed.
(c) Stationary
\(|\langle S_z/|\mathbf S|\rangle|\) versus
\(\lambda_{\rm relax}\) at the same
\((P/P_{\rm th},\gamma_{a,E})\) coordinates.
(d) Smooth-ramp acoustic response \(A_z\) at
\(\beta_{h,E}=0.62~\upmu\mathrm{eV}\) for several turn-on times.
The vertical dotted line marks the resonant drive energy
\(E_0\omega=9.13~\upmu\mathrm{eV}\). The inset shows the pseudospin trajectory near the switching event for the slowest ramp.}
    \label{fig:lifetime_anisotropy}
\end{figure}

We next include polarization-dependent losses of the two linearly polarized
modes, as in Ref.~\cite{Ohadi2015}. We keep \(\Delta=0\), so that no external
magnetic field explicitly favors either circular polarization, but allow
\(\gamma_a\neq0\) in Eq.~\eqref{Sdyn}. With the sign convention used here,
positive \(\gamma_a\) favors states with negative \(S_x\). In combination with
the static in-plane splitting, spin relaxation, and the nonlinear self-induced
field, this anisotropy can destabilize the purely in-plane polarization state
and produce a dissipative symmetry-breaking bifurcation with a finite
out-of-plane pseudospin component. The stationary algebraic equations underlying this bifurcation and the
symmetry-related \(S_z\to -S_z\) branches are summarized in Supplementary
Note~5 of the Supplemental Material.

The stationary bifurcation is summarized in
Fig.~\ref{fig:lifetime_anisotropy}. In panels (a) and (b), the color scale
shows \(|\langle S_z/|\mathbf S|\rangle|\) in the
\((P/P_{\rm th},\gamma_{a,E})\) plane in the absence of acoustic driving. A broad region with finite normalized circular-polarization component,
\(|\langle S_z/|\mathbf S|\rangle|\), appears even though \(\Delta=0\). Thus the stationary condensate develops elliptical polarization without any
Zeeman bias, through a dissipative symmetry-breaking mechanism. The comparison between panels (a) and (b) shows
that spin relaxation substantially reshapes the bifurcated region and can
strongly restrict the parameter range where stable out-of-plane states survive.

\begin{figure}[t]
    \centering
    \includegraphics[width=\columnwidth]{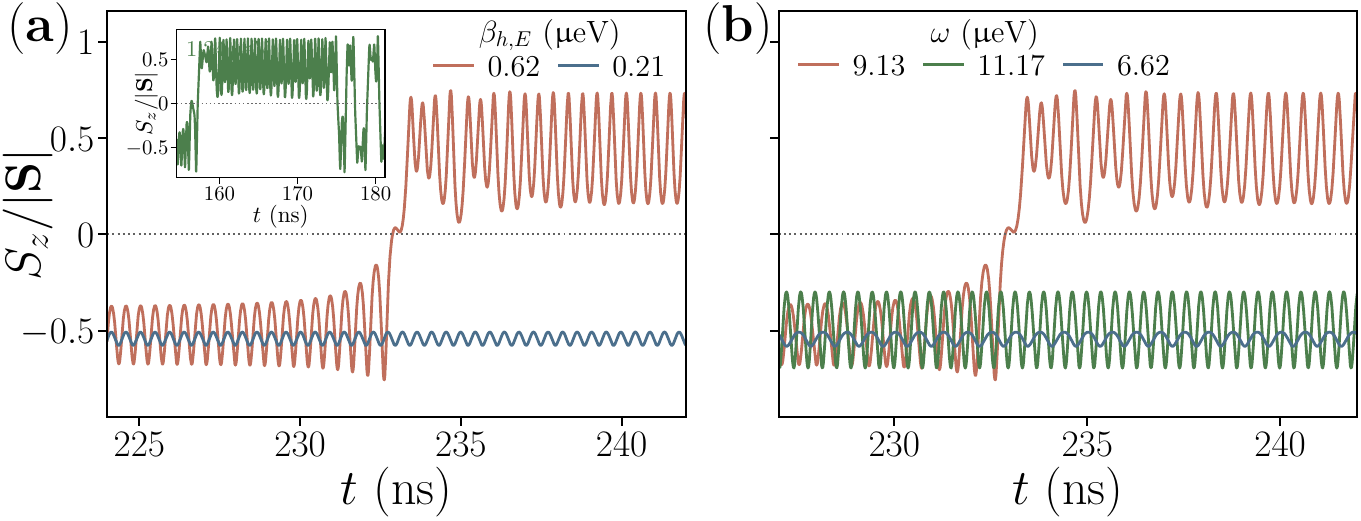}
    \caption{\textbf{Time-domain acoustic switching in the bifurcated regime.}
The working point is the one marked in
Fig.~\ref{fig:lifetime_anisotropy}:
\(P/P_{\rm th}=1.30\),
\(\gamma_{a,E}=3.03~\upmu\mathrm{eV}\), and
\(\lambda_{\rm relax}=0.05\), with smooth turn-on time
\(\tau_{\rm on}=30.56~{\rm ns}\).
All driven traces start from the relaxed undriven stationary state
selected by the preparatory protocol described in
Fig.~\ref{fig:lifetime_anisotropy}:
$(S_x,S_y,S_z,n_R)(0)\simeq(-1.610,0.162,-1.046,0.877),
\ \frac{\mathbf S(0)}{|\mathbf S(0)|}\simeq(-0.836,0.084,-0.543).$
(a) Driven traces of \(S_z/|\mathbf S|\) at the resonant drive energy
\(E_0\omega=9.13~\upmu\mathrm{eV}\) for
\(\beta_{h,E}=0.62~\upmu\mathrm{eV}\) and
\(0.21~\upmu\mathrm{eV}\). The stronger drive produces switching between the
two dissipative branches, whereas the weaker drive remains near the initial
branch. The inset shows the switching interval for the larger test amplitude
\(\beta_{h,E}=1.24~\upmu\mathrm{eV}\).
(b) Driven traces at fixed
\(\beta_{h,E}=0.62~\upmu\mathrm{eV}\) for
\(E_0\omega=9.13,11.17,6.62~\upmu\mathrm{eV}\). Switching occurs near the resonant drive energy, while off-resonant traces remain close to the original branch.}
    \label{fig:bifurcated_switching_traces}
\end{figure}

The white marker in Fig.~\ref{fig:lifetime_anisotropy}(b) denotes the working
point used for the acoustic switching simulations:
\(P/P_{\rm th}=1.30\),
\(\gamma_{a,E}=E_0\gamma_a=3.03~\upmu\mathrm{eV}\), and
\(\lambda_{\rm relax}=0.05\). For the bifurcated-state simulations shown in
Figs.~\ref{fig:lifetime_anisotropy} and
\ref{fig:bifurcated_switching_traces}, the system was first relaxed without
acoustic drive from a branch-selecting initial condition with
\(S_z/|\mathbf S|\simeq0.24\), as displayed in the inset of
Fig.~\ref{fig:lifetime_anisotropy}(b). The undriven relaxation selects a
stationary branch with \(S_z/|\mathbf S|\simeq-0.543\); the opposite branch is
obtained by reversing the sign of the initial circular-polarization seed. The
acoustic drive in the switching simulations is then switched on starting from
this relaxed undriven stationary state. Panel (c) follows the same working point
while varying \(\lambda_{\rm relax}\). Increasing spin relaxation suppresses the stationary circular-polarization
component because it damps transverse pseudospin motion and drives the spin more efficiently toward the stable in-plane eigenaxis set mainly by the static linear-polarization splitting.

We then apply the same \(Y\)-polarized acoustic geometry as in the previous
subsection. Since the unforced stationary state already possesses a finite
out-of-plane pseudospin component, the relevant driven quantity is now \(A_z\).
Figure~\ref{fig:lifetime_anisotropy}(d) shows a pronounced resonance near
\(E_0\omega=9.13~\upmu\mathrm{eV}\), substantially below the bare
linear-polarization splitting
\(\delta_E=16.54~\upmu\mathrm{eV}\). We therefore interpret this resonance as
the excitation of a softened polarization mode associated with the dissipatively
bifurcated state, rather than as simple precession around the bare in-plane
field. The dependence on the turn-on time shows that a slower ramp suppresses
transient oscillations and allows the driven trajectory to approach the
nonlinear switching regime.

The corresponding time-domain dynamics is shown in
Fig.~\ref{fig:bifurcated_switching_traces}. At the resonant drive energy, a sufficiently strong resonant acoustic drive
produces a transition from the initial out-of-plane branch to the branch with
opposite-sign circular polarization, whereas a weaker drive remains trapped near
the original stationary state. The inset in
Fig.~\ref{fig:bifurcated_switching_traces}(a) resolves the switching interval
for a larger test amplitude and shows the rapid pseudospin transfer between the
two dissipatively selected branches.

Panel~\ref{fig:bifurcated_switching_traces}(b) demonstrates that this transfer
is strongly resonant. Near
\(E_0\omega=9.13~\upmu\mathrm{eV}\), the acoustic field excites sufficiently large pseudospin motion to push the
trajectory from the basin of the initial branch into the basin of the opposite
branch. Off resonance, the pseudospin performs only small oscillations
around the initial branch and no switching occurs. Thus, in the bifurcated
regime, the acoustic wave can do more than induce weak polarization oscillations:
it can trigger a controlled transition between metastable dissipative
polarization states. We emphasize that the simulations demonstrate a single
acoustic switching event rather than a periodic back-and-forth switching cycle.
The switching is therefore controlled by both resonance and amplitude: the drive
must be close enough to the softened polarization mode and strong enough to move
the trajectory into the basin of the opposite branch.

\subsection{Zeeman tuning of the acoustic resonance}

\begin{figure}[t]
    \centering
    \includegraphics[width=\columnwidth]{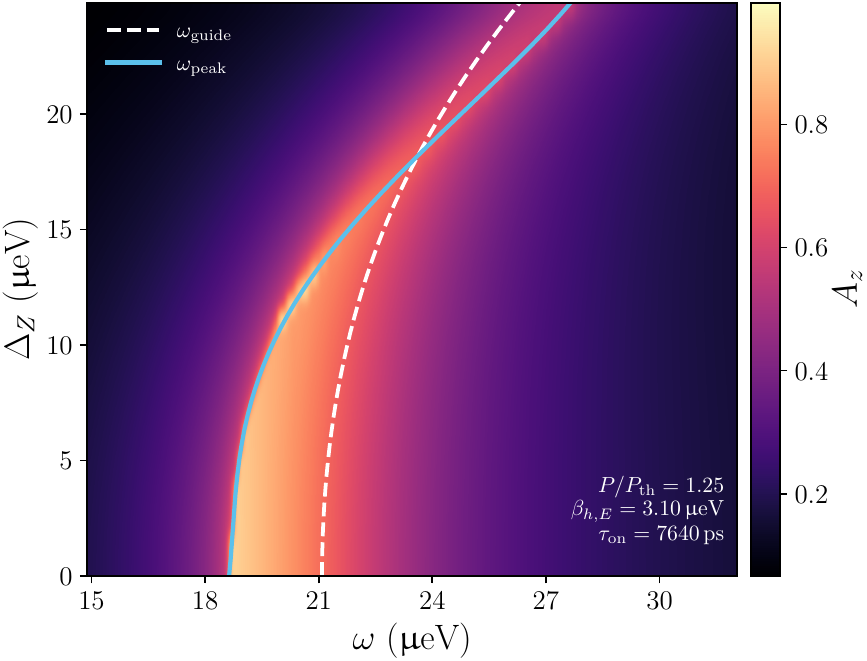}
    \caption{\textbf{Zeeman-tuned acoustic resonance map.}
The minimal pumped--dissipative model is used with \(\gamma_a=0\),
\(P/P_{\rm th}=1.25\), \(\alpha_s=0.5\),
\(\lambda_{\rm relax}=0.05\), and acoustic amplitude
\(\beta_{h,E}=3.10~\upmu\mathrm{eV}\). The drive is switched on
smoothly with \(\tau_{\rm on}=7640~\mathrm{ps}\).
For each value of \(\Delta\), the system is first relaxed at
\(\beta_h=0\) from $(S_x,S_y,S_z,n_R)(0)=(1,0,0,1),$
and the acoustic drive is then applied starting from the resulting
\(\Delta\)-dependent undriven stationary state.
The color scale shows the circular-polarization response amplitude \(A_z\) in
the \((\omega,\Delta_Z)\) plane, with \(\omega\) displayed through the acoustic
drive energy \(E_0\omega=\hbar\Omega\). The cyan curve is the numerically
extracted resonance ridge \(\omega_{\rm peak}\), while the white dashed
curve is the small-oscillation guide \(\omega_{\rm guide}\). Increasing
\(\Delta_Z=E_0\Delta\) shifts the acoustic resonance to higher drive energy.}
    \label{fig:zeeman_tuned}
\end{figure}

Finally, we restore a finite Zeeman splitting while keeping
\(\gamma_a=0\). In Eq.~\eqref{eq:static_field}, this adds a controllable
out-of-plane contribution to the effective magnetic field, $B_z=\alpha_s S_z-\Delta .$
This case is qualitatively different from the
lifetime-anisotropy-induced bifurcation discussed above: the out-of-plane
pseudospin component is now selected explicitly by the external magnetic field,
rather than by spontaneous dissipative symmetry breaking.

For each value of \(\Delta\), the system is first relaxed without acoustic
drive and then driven by the same transverse acoustic field, with
\(\phi=\pi/4\). The resulting response map is shown in
Fig.~\ref{fig:zeeman_tuned}. Increasing the physical Zeeman splitting
\(\Delta_Z=E_0\Delta\) shifts the resonance ridge to higher acoustic drive
energies. Thus the external magnetic field provides a conservative and
experimentally accessible tuning knob for the acoustic spin resonance.

The white dashed curve in Fig.~\ref{fig:zeeman_tuned} is a small-oscillation
guide. It is obtained by linearizing the conservative part of the pseudospin dynamics
around the undriven pumped--dissipative stationary state
\(\mathbf S_{\rm st}\). With
\[
B_z^{\rm st}=\alpha_s S_z^{\rm st}-\Delta ,
\]
the corresponding linearized matrix reads
\[
J=
\begin{pmatrix}
0 & -B_z^{\rm st} & -\alpha_s S_y^{\rm st}\\
B_z^{\rm st} & 0 & \delta+\alpha_s S_x^{\rm st}\\
0 & -\delta & 0
\end{pmatrix}.
\]
The nonzero eigenvalue pair of \(J\) is written as
\(\lambda_\pm=\pm i\omega_{\rm guide}\), which defines the guide frequency
\(\omega_{\rm guide}\). The derivation of this guide frequency and its conversion to the displayed
energy scale are given in Supplementary Note~6 of the Supplemental Material. This small-oscillation estimate captures the overall
upward trend of the resonance ridge but is not expected to coincide exactly
with the numerical maximum, because the latter is extracted from
finite-amplitude driven dynamics in a nonlinear pumped--dissipative system.

\section{Conclusions}

We have developed a minimal theory of acoustic spin resonance in a spatially
homogeneous spinor polariton condensate. A longitudinal acoustic wave produces a
time-periodic strain-induced effective magnetic field acting on the condensate
pseudospin. When the acoustic propagation direction is chosen so that this field
is transverse to the static in-plane splitting, the acoustic drive resonantly
excites polarization oscillations.

In the conservative limit, spin-dependent interactions renormalize the
precession frequency, shift the resonance, and produce nonlinear line shapes. In
the pumped--dissipative regime, reservoir dynamics and spin relaxation stabilize
nonlinear driven states, turning the resonance into a history-dependent response
with amplitude hysteresis. We have further shown that lifetime anisotropy can
create a bifurcated stationary state with a finite circular-polarization
component. In this regime, a resonant acoustic drive can trigger switching
between the corresponding out-of-plane polarization branches. Finally, a Zeeman
splitting provides an independent conservative tuning knob: by tilting the
stationary pseudospin and changing the effective precession frequency, it shifts
the acoustic-resonance ridge in a controllable way.

These results establish coherent acoustic driving as a flexible mechanism for
resonant, nonlinear, and switchable control of polariton pseudospin dynamics.
They also suggest that strain-based acoustic fields can be used as a dynamical
control resource in spinor polariton systems, complementing optical and magnetic
methods of polarization manipulation.

\begin{acknowledgments}
A.K.\ was supported by the Icelandic Research Fund (Grant No.~2410550).\end{acknowledgments}


\bibliography{main}

@article{lanzillotti2007coherent,
  title={Coherent generation of acoustic phonons in an optical microcavity},
  author={Lanzillotti-Kimura, ND and Fainstein, A and Huynh, Agn{\`e}s and Perrin, Bernard and Jusserand, Bernard and Miard, A and Lema{\^\i}tre, A},
  journal={Physical review letters},
  volume={99},
  number={21},
  pages={217405},
  year={2007},
  publisher={APS}
}

@article{vishnevsky2011coherent,
  title = {Coherent interactions between phonons and exciton or exciton-polariton condensates},
  author = {Vishnevsky, D. V. and Solnyshkov, D. D. and Malpuech, G. and Gippius, N. A. and Shelykh, I. A.},
  journal = {Phys. Rev. B},
  volume = {84},
  pages = {035312},
  year = {2011},
  doi = {10.1103/PhysRevB.84.035312},
  url = {https://doi.org/10.1103/PhysRevB.84.035312},
  publisher = {American Physical Society}
}

@article{yulin2019resonant,
  title={Resonant excitation of acoustic waves in one-dimensional exciton-polariton systems},
  author={Yulin, AV and Kozin, VK and Nalitov, AV and Shelykh, IA},
  journal={Physical Review A},
  volume={100},
  number={4},
  pages={043610},
  year={2019},
  publisher={APS}
}

@article{carraro2024solid,
  title={Solid-state continuous time crystal in a polariton condensate with a built-in mechanical clock},
  author={Carraro-Haddad, Ignacio and Chafatinos, Dimitri Lisandro and Kuznetsov, AS and Papuccio-Fern{\'a}ndez, Ignacio Agust{\'\i}n and Reynoso, Andres Alejandro and Bruchhausen, A and Biermann, K and Santos, PV and Usaj, Gonzalo and Fainstein, Alejandro},
  journal={Science},
  volume={384},
  number={6699},
  pages={995--1000},
  year={2024},
  publisher={American Association for the Advancement of Science}
}

@article{Chafatinos2020,
  title = {Polariton-driven phonon laser},
  author = {Chafatinos, D. L. and Kuznetsov, A. S. and Anguiano, S.
  and Bruchhausen, A. E. and Reynoso, A. A. and Biermann, K.
  and Santos, P. V. and Fainstein, A.},
  journal = {Nat. Commun.},
  volume = {11},
  pages = {4552},
  year = {2020},
  doi = {10.1038/s41467-020-18358-z}
}

@article{Ivanov2003,
  author = {Ivanov, A. L. and Littlewood, P. B.},
  title = {Resonant acousto-optics of microcavity polaritons},
  journal = {Semicond. Sci. Technol.},
  volume = {18},
  number = {10},
  pages = {S428--S434},
  year = {2003},
  doi = {10.1088/0268-1242/18/10/318}
}

@article{CerdaMendez2010,
  author = {Cerda-M{\'e}ndez, E. A. and Krizhanovskii, D. N. and Wouters, M. and Bradley, R. and Biermann, K. and Guda, K. and Hey, R. and Santos, P. V. and Sarkar, D. and Skolnick, M. S.},
  title = {Polariton Condensation in Dynamic Acoustic Lattices},
  journal = {Phys. Rev. Lett.},
  volume = {105},
  pages = {116402},
  year = {2010},
  doi = {10.1103/PhysRevLett.105.116402}
}

@article{Kuznetsov2019,
  author = {Kuznetsov, Alexander S. and Biermann, Klaus and Santos, Paulo V.},
  title = {Dynamic acousto-optical control of confined polariton condensates: From single traps to coupled lattices},
  journal = {Phys. Rev. Research},
  volume = {1},
  pages = {023030},
  year = {2019},
  doi = {10.1103/PhysRevResearch.1.023030}
}

@article{RamosPerez2024,
  author = {Ramos-P{\'e}rez, I. A. and Carraro-Haddad, I. and Fainstein, F. and Chafatinos, D. L. and Usaj, G. and Mindlin, G. B. and Fainstein, A. and Reynoso, A. A.},
  title = {Theory of optomechanical locking in driven-dissipative coupled polariton condensates},
  journal = {Phys. Rev. B},
  volume = {109},
  pages = {165305},
  year = {2024},
  doi = {10.1103/PhysRevB.109.165305}
}

@article{cho2005bragg,
  author = {Cho, Kikuo and Okumoto, Kazunori and Nikolaev, N. I. and Ivanov, A. L.},
  title = {Bragg Diffraction of Microcavity Polaritons by a Surface Acoustic Wave},
  journal = {Phys. Rev. Lett.},
  volume = {94},
  pages = {226406},
  year = {2005},
  doi = {10.1103/PhysRevLett.94.226406},
  url = {https://doi.org/10.1103/PhysRevLett.94.226406}
}

@article{Deng2010,
  author = {Deng, Hui and Haug, Hartmut and Yamamoto, Yoshihisa},
  title = {Exciton-polariton {Bose--Einstein} condensation},
  journal = {Rev. Mod. Phys.},
  volume = {82},
  number = {2},
  pages = {1489--1537},
  year = {2010},
  doi = {10.1103/RevModPhys.82.1489},
  publisher = {American Physical Society}
}

@article{Balili2010,
  title = {Huge splitting of polariton states in microcavities under stress},
  author = {Balili, R. and Nelsen, B. and Snoke, D. W. and Reid, R. H. and Pfeiffer, L. and West, K.},
  journal = {Phys. Rev. B},
  volume = {81},
  issue = {12},
  pages = {125311},
  numpages = {8},
  year = {2010},
  month = {Mar},
  publisher = {American Physical Society},
  doi = {10.1103/PhysRevB.81.125311},
  url = {https://link.aps.org/doi/10.1103/PhysRevB.81.125311}
}

@article{Klopotowski2006,
  title = {Optical anisotropy and pinning of the linear polarization of light in semiconductor microcavities},
  author = {Klopotowski, L. and Martin, M. D. and  Amo, A. and Vina, L. and Shelykh, I. A. and  Glazov, M. M. and Malpuech, G. and Kavokin, A. V. and Andr{\'e}, R.},
  journal = {Solid State Commun.},
  volume = {139},
  pages = {511},
  numpages = {8},
  year = {2006},
  doi = {10.1016/j.ssc.2006.07.016},
  url = {https://www.sciencedirect.com/science/article/pii/S0038109806006211?via%3Dihub}
}

@article{Saltykova2025spin,
  title = {Spin relaxation in a polariton fluid: Quantum hydrodynamic approach},
  author = {Saltykova, D. A. and Yulin, A. V. and Shelykh, I. A.},
  journal = {Phys. Rev. B},
  volume = {113},
  pages = {134513},
  year = {2026},
  doi = {10.1103/PhysRevB.113.134513}
}

@article{Kasprzak2007,
  title = {Build up and pinning of linear polarization in the Bose condensates of exciton polaritons},
  author = {Kasprzak, J. and Andr\'e, R. and Dang, Le Si and Shelykh, I. A. and Kavokin, A. V. and Rubo, Yuri G. and Kavokin, K. V. and Malpuech, G.},
  journal = {Phys. Rev. B},
  volume = {75},
  issue = {4},
  pages = {045326},
  numpages = {5},
  year = {2007},
  month = {Jan},
  publisher = {American Physical Society},
  doi = {10.1103/PhysRevB.75.045326},
  url = {https://link.aps.org/doi/10.1103/PhysRevB.75.045326}
}

@article{Ohadi2015,
  title = {Spontaneous Spin Bifurcations and Ferromagnetic Phase Transitions in a Spinor Exciton-Polariton Condensate},
  author = {Ohadi, H. and Dreismann, A. and Rubo, Y. G. and Pinsker, F. and del Valle-Inclan Redondo, Y. and Tsintzos, S. I. and Hatzopoulos, Z. and Savvidis, P. G. and Baumberg, J. J.},
  journal = {Phys. Rev. X},
  volume = {5},
  issue = {3},
  pages = {031002},
  numpages = {18},
  year = {2015},
  month = {Jul},
  publisher = {American Physical Society},
  doi = {10.1103/PhysRevX.5.031002},
  url = {https://link.aps.org/doi/10.1103/PhysRevX.5.031002}
}

@article{Baumberg2008,
  title = {Spontaneous Polarization Buildup in a Room-Temperature Polariton Laser},
  author = {Baumberg, J. J. and Kavokin, A. V. and Christopoulos, S. and Grundy, A. J. D. and Butt\'e, R. and Christmann, G. and Solnyshkov, D. D. and Malpuech, G. and Baldassarri H\"oger von H\"ogersthal, G. and Feltin, E. and Carlin, J.-F. and Grandjean, N.},
  journal = {Phys. Rev. Lett.},
  volume = {101},
  issue = {13},
  pages = {136409},
  numpages = {4},
  year = {2008},
  month = {Sep},
  publisher = {American Physical Society},
  doi = {10.1103/PhysRevLett.101.136409},
  url = {https://link.aps.org/doi/10.1103/PhysRevLett.101.136409}
}

@article{ShelykhSILP,
  title = {Spin dynamics of interacting exciton polaritons in microcavities},
  author = {Shelykh, I. and Malpuech, G. and Kavokin, K. V. and Kavokin, A. V. and Bigenwald, P.},
  journal = {Phys. Rev. B},
  volume = {70},
  issue = {11},
  pages = {115301},
  numpages = {12},
  year = {2004},
  month = {Sep},
  publisher = {American Physical Society},
  doi = {10.1103/PhysRevB.70.115301},
  url = {https://link.aps.org/doi/10.1103/PhysRevB.70.115301}
}

@article{Carusotto2013,
  title = {Quantum fluids of light},
  author = {Carusotto, Iacopo and Ciuti, Cristiano},
  journal = {Rev. Mod. Phys.},
  volume = {85},
  issue = {1},
  pages = {299--366},
  numpages = {0},
  year = {2013},
  month = {Feb},
  publisher = {American Physical Society},
  doi = {10.1103/RevModPhys.85.299},
  url = {https://link.aps.org/doi/10.1103/RevModPhys.85.299}
}

@article{Ballarini2017,
  title = {Macroscopic Two-Dimensional Polariton Condensates},
  author = {Ballarini, Dario and Caputo, Davide and Mu\~noz, Carlos S\'anchez and De Giorgi, Milena and Dominici, Lorenzo and Szyma\ifmmode \acute{n}\else \'{n}\fi{}ska, Marzena H. and West, Kenneth and Pfeiffer, Loren N. and Gigli, Giuseppe and Laussy, Fabrice P. and Sanvitto, Daniele},
  journal = {Phys. Rev. Lett.},
  volume = {118},
  issue = {21},
  pages = {215301},
  numpages = {5},
  year = {2017},
  month = {May},
  publisher = {American Physical Society},
  doi = {10.1103/PhysRevLett.118.215301},
  url = {https://link.aps.org/doi/10.1103/PhysRevLett.118.215301}
}

@article{Kasprzak2006,
  title = {Bose–Einstein condensation of exciton polaritons},
  author = {Kasprzak, J. and Richard, M. and Kundermann, S. and Baas, A. and Jeambrun, P. and Keeling, J. M. J. and Marchetti, F. M. and Szyma{\'n}ska, M. H. and Andr{\'e}, R. and Staehli, J. M. and Savona, V. and Littlewood, P. B. and Deveaud, B. and Dang, Le Si},
  journal = {Nature},
  volume = {443},
  pages = {409--414},
  numpages = {0},
  year = {2006},
  doi = {10.1038/nature05131},
  url = {https://www.nature.com/articles/nature05131}
}

@article{ShelykhReview,
  title = {Polariton polarization-sensitive phenomena in planar semiconductor microcavities},
  author = {Shelykh, I. A. and Kavokin, A. V. and Rubo, Yu. G and Liew, T. C. H. and Malpuech, G.},
  journal = {Semicond. Sci. Technol.},
  volume = {25},
  pages = {013001},
  year = {2010},
  doi = {10.1088/0268-1242/25/1/013001},
  url = {https://iopscience.iop.org/article/10.1088/0268-1242/25/1/013001?pageTitle=IOPscience}
}

@article{Ciuti1998,
  title = {Role of the exchange of carriers in elastic exciton-exciton scattering in quantum wells},
  author = {Ciuti, C. and Savona, V. and Piermarocchi, C. and Quattropani, A. and Schwendimann, P.},
  journal = {Phys. Rev. B},
  volume = {58},
  issue = {12},
  pages = {7926--7933},
  numpages = {0},
  year = {1998},
  month = {Sep},
  publisher = {American Physical Society},
  doi = {10.1103/PhysRevB.58.7926},
  url = {https://link.aps.org/doi/10.1103/PhysRevB.58.7926}
}

@article{Glazov2009,
  title = {Polariton-polariton scattering in microcavities: A microscopic theory},
  author = {Glazov, M. M. and Ouerdane, H. and Pilozzi, L. and Malpuech, G. and Kavokin, A. V. and D'Andrea, A.},
  journal = {Phys. Rev. B},
  volume = {80},
  issue = {15},
  pages = {155306},
  numpages = {14},
  year = {2009},
  month = {Oct},
  publisher = {American Physical Society},
  doi = {10.1103/PhysRevB.80.155306},
  url = {https://link.aps.org/doi/10.1103/PhysRevB.80.155306}
}
 
\end{document}


\title{Supplementary Materials: Acoustic spin resonance in polariton condensates}

\author{D. A. Saltykova}
\affiliation{Department of Physics, ITMO University, Saint Petersburg 197101, Russia}

\author{A. Kudlis}
\affiliation{Science Institute, University of Iceland, Dunhagi 3, IS-107, Reykjavik, Iceland}

\author{A. V. Yulin}
\affiliation{Department of Physics, ITMO University, Saint Petersburg 197101, Russia}

\author{I. A. Shelykh}
\affiliation{Science Institute, University of Iceland, Dunhagi 3, IS-107, Reykjavik, Iceland}

\maketitle
\onecolumngrid
\tableofcontents

\section*{Supplementary Note 1: Derivation of the effective pseudospin equations}

In this Supplementary Note we derive the effective pseudospin equations used in
the main text. Unless stated otherwise, all quantities in this note are written
in the dimensionless units defined in Supplementary Note~2.

We consider a spatially homogeneous spinor polariton condensate described by the
order parameter
\begin{equation}
\Psi(t)=
\begin{pmatrix}
\psi_+(t)\\
\psi_-(t)
\end{pmatrix},
\end{equation}
where \(\psi_\pm\) are the amplitudes of the two circularly polarized
condensate components. The condensate density and pseudospin are defined as
\begin{equation}
n=\Psi^\dagger\Psi=|\psi_+|^2+|\psi_-|^2,
\qquad
S_i=\Psi^\dagger\sigma_i\Psi,
\end{equation}
where \(\sigma_i\) are the Pauli matrices. Explicitly,
\begin{equation}
S_x=2\mathrm{Re}(\psi_+^*\psi_-),\qquad
S_y=2\mathrm{Im}(\psi_+^*\psi_-),\qquad
S_z=|\psi_+|^2-|\psi_-|^2.
\end{equation}
Therefore
\begin{equation}
|\mathbf S|^2=S_x^2+S_y^2+S_z^2
=
\left(|\psi_+|^2+|\psi_-|^2\right)^2=n^2,
\end{equation}
so that the pseudospin length is equal to the condensate density,
\begin{equation}
|\mathbf S|=n.
\end{equation}

The coherent part of the condensate dynamics is governed by the spinor equation
\begin{equation}
i\partial_t\Psi=\hat H_{\mathrm{eff}}\Psi,
\label{eq:supp_spinor_eq}
\end{equation}
with
\begin{equation}
\hat H_{\mathrm{eff}}
=
\frac12
\left[
\mathbf B_0(\mathbf S)+\mathbf B_a(t)
\right]\cdot\boldsymbol{\sigma}.
\label{eq:supp_Heff}
\end{equation}

The stationary effective magnetic field is taken in the form
\begin{equation}
\mathbf B_0(\mathbf S)
=
-\delta\,\mathbf e_x
+
\left(\alpha_s S_z-\Delta\right)\mathbf e_z.
\label{eq:supp_B0}
\end{equation}
Here \(\delta\) is the static splitting between two linearly polarized cavity
modes caused by intrinsic birefringence, \(\Delta\) is the Zeeman splitting
between the two circularly polarized states, and
\(\alpha_s S_z\) is the self-induced circular-polarization splitting due to
spin-anisotropic polariton--polariton interactions. The common scalar
interaction energy, proportional to the identity matrix in spin space, is
omitted because it does not affect the pseudospin dynamics.

The acoustic wave produces a time-dependent in-plane effective field. In the
most general form used in the simulations we write it as
\begin{equation}
\mathbf B_a(t)
=
\beta_h(t)
\left(
\mathbf e_x\cos 2\phi+\mathbf e_y\sin 2\phi
\right)
\sin\Theta(t),
\qquad
\dot\Theta(t)=\omega(t),
\label{eq:supp_Ba_general}
\end{equation}
where \(\phi\) is the propagation angle of the acoustic wave. The angular
factor \(2\phi\) follows from the transformation of the strain-induced linear
birefringence under an in-plane rotation. For a fixed frequency and fixed
acoustic amplitude, Eq.~\eqref{eq:supp_Ba_general} reduces to
\begin{equation}
\mathbf B_a(t)
=
\beta_h
\left(
\mathbf e_x\cos 2\phi+\mathbf e_y\sin 2\phi
\right)
\sin\omega t .
\label{eq:supp_Ba_fixed}
\end{equation}
Thus \(\beta_h\) is the amplitude of the strain-induced linear-polarization
splitting.

We now derive the pseudospin precession equation. Differentiating
\(S_i=\Psi^\dagger\sigma_i\Psi\) with respect to time and using
Eq.~\eqref{eq:supp_spinor_eq}, one obtains
\begin{equation}
\partial_t S_i
=
i\Psi^\dagger
\left[
\hat H_{\mathrm{eff}},\sigma_i
\right]
\Psi .
\end{equation}
Using
\begin{equation}
[\sigma_j,\sigma_i]=2i\varepsilon_{jik}\sigma_k,
\end{equation}
we find
\begin{align}
\partial_t S_i
&=
\frac{i}{2}
B_j
\Psi^\dagger
[\sigma_j,\sigma_i]
\Psi=
\frac{i}{2}
B_j
2i\varepsilon_{jik}
S_k
\nonumber\\
&=
-B_j\varepsilon_{jik}S_k
=
\varepsilon_{ijk}B_jS_k.
\end{align}
Therefore the coherent pseudospin dynamics is
\begin{equation}
\partial_t\mathbf S
=
\left[
\mathbf B_0(\mathbf S)+\mathbf B_a(t)
\right]\times \mathbf S.
\label{eq:supp_coherent_spin}
\end{equation}
This sign corresponds to the convention
\(\hat H_{\mathrm{eff}}=\tfrac12\mathbf B\cdot\boldsymbol{\sigma}\).

We now include gain and loss. The reservoir is assumed to be unpolarized, so
stimulated scattering from the reservoir acts equally on both spin components.
At the spinor level this contribution can be written as
\begin{equation}
\left.\partial_t\Psi\right|_{\mathrm{gain}}
=
\frac12 Rn_R\Psi,
\end{equation}
where \(R\) is the reservoir-to-condensate scattering rate and \(n_R\) is the
reservoir density. The factor \(1/2\) appears because \(Rn_R\) is the growth
rate of the condensate density, whereas the spinor amplitude grows with half
that rate.

The polarization-independent radiative decay is similarly described by
\begin{equation}
\left.\partial_t\Psi\right|_{\mathrm{loss},0}
=
-\frac{\gamma}{2}\Psi.
\end{equation}
Together, the spin-independent gain and loss give
\begin{equation}
\left.\partial_t\Psi\right|_{\mathrm{iso}}
=
\frac12(Rn_R-\gamma)\Psi.
\label{eq:supp_iso_gain_loss_spinor}
\end{equation}
They lead to the pseudospin contribution
\begin{equation}
\left.\partial_t\mathbf S\right|_{\mathrm{iso}}
=
(Rn_R-\gamma)\mathbf S.
\label{eq:supp_iso_gain_loss_spin}
\end{equation}

We next derive the anisotropic loss term used in the main text. Following the
phenomenological pseudospin description of polarization-dependent lifetimes used
in Ref.~\cite{Ohadi2015}, we assume that the two orthogonal linearly polarized
modes have different decay rates. Let
\begin{equation}
\psi_X=\frac{\psi_++\psi_-}{\sqrt2},
\qquad
\psi_Y=\frac{i(\psi_+-\psi_-)}{\sqrt2}
\end{equation}
be the two linearly polarized components. With this convention,
\begin{equation}
S_x=|\psi_X|^2-|\psi_Y|^2.
\label{eq:supp_Sx_XY}
\end{equation}
Assume that these two linear polarizations have different decay rates,
\begin{equation}
\gamma_X=\gamma+\gamma_a,
\qquad
\gamma_Y=\gamma-\gamma_a.
\label{eq:supp_gamma_XY}
\end{equation}
In the \((X,Y)\) basis the decay matrix is therefore diagonal:
\begin{equation}
\hat\Gamma_{XY}
=
\begin{pmatrix}
\gamma_X & 0\\
0 & \gamma_Y
\end{pmatrix}
=
\gamma\,\mathbb 1+\gamma_a\,\sigma_z^{(XY)}.
\end{equation}
Transforming back to the circular basis, this anisotropic part becomes
proportional to \(\sigma_x\). Hence the gain--loss contribution to the spinor
equation in the circular basis is
\begin{equation}
\left.\partial_t\Psi\right|_{\mathrm{gain/loss}}
=
\frac12
\left[
(Rn_R-\gamma)\mathbb 1-\gamma_a\sigma_x
\right]\Psi.
\label{eq:supp_gain_loss_spinor_full}
\end{equation}

Using Eq.~\eqref{eq:supp_gain_loss_spinor_full}, the corresponding contribution
to \(S_i=\Psi^\dagger\sigma_i\Psi\) is
\begin{align}
\left.\partial_t S_i\right|_{\mathrm{gain/loss}}
&=
(Rn_R-\gamma)S_i
-\frac{\gamma_a}{2}
\Psi^\dagger
\left(
\sigma_x\sigma_i+\sigma_i\sigma_x
\right)
\Psi
\nonumber\\
&=
(Rn_R-\gamma)S_i
-\frac{\gamma_a}{2}
\Psi^\dagger
\{\sigma_x,\sigma_i\}
\Psi.
\end{align}
Since
\begin{equation}
\{\sigma_x,\sigma_i\}=2\delta_{xi}\mathbb 1,
\end{equation}
we obtain
\begin{equation}
\left.\partial_t S_i\right|_{\mathrm{gain/loss}}
=
(Rn_R-\gamma)S_i-\gamma_a n\,\delta_{xi}.
\end{equation}
In vector form,
\begin{equation}
\left.\partial_t\mathbf S\right|_{\mathrm{gain/loss}}
=
(Rn_R-\gamma)\mathbf S
-
\gamma_a n\,\mathbf e_x.
\label{eq:supp_ohadi_term}
\end{equation}
Since \(n=|\mathbf S|\), this is exactly the term used in the main text:
\begin{equation}
\left.\partial_t\mathbf S\right|_{\mathrm{gain/loss}}
=
(Rn_R-\gamma)\mathbf S
-
\gamma_a |\mathbf S|\,\mathbf e_x.
\label{eq:supp_ohadi_term_absS}
\end{equation}
The sign of \(\gamma_a\) depends on which of the two
linear polarizations has the shorter lifetime. With the convention
\eqref{eq:supp_gamma_XY}, positive \(\gamma_a\) means that the \(X\)-polarized
mode decays faster than the \(Y\)-polarized one.

The same derivation gives the corresponding density equation. From
\(n=\Psi^\dagger\Psi\), Eq.~\eqref{eq:supp_gain_loss_spinor_full} yields
\begin{align}
\left.\partial_t n\right|_{\mathrm{gain/loss}}
&=
(Rn_R-\gamma)n
-
\gamma_a\Psi^\dagger\sigma_x\Psi
\nonumber\\
&=
(Rn_R-\gamma)n-\gamma_a S_x.
\label{eq:supp_density_gain_loss}
\end{align}
This density equation is not written separately in the main text because
\(n=|\mathbf S|\), and therefore it follows directly from the pseudospin
equation. Thus the anisotropic decay affects not only the polarization dynamics
but also the total condensate density, depending on the instantaneous linear
polarization of the condensate.

The reservoir is assumed to be scalar and unpolarized. It is described by
\begin{equation}
\partial_t n_R
=
P-\gamma_R n_R-Rn_R n.
\label{eq:supp_reservoir_eq_n}
\end{equation}
Using \(n=|\mathbf S|\), this becomes
\begin{equation}
\partial_t n_R
=
P-\gamma_R n_R-Rn_R|\mathbf S|.
\label{eq:supp_reservoir_eq_S}
\end{equation}

Finally, following the hydrodynamic spin-relaxation phenomenology introduced in
Ref.~\cite{Saltykova2025spin}, we add spin relaxation in the form of a
Gilbert-type term,
\begin{equation}
\left.\partial_t\mathbf S\right|_{\mathrm{rel}}
=
-\lambda\,
\mathbf S\times
\left[
\mathbf B_0(\mathbf S)\times\mathbf S
\right].
\label{eq:supp_gilbert}
\end{equation}
Here \(\lambda\) is the spin-relaxation coefficient. In this work the relaxation
field is chosen to be the stationary field \(\mathbf B_0(\mathbf S)\). This
choice corresponds to relaxation towards the spin eigenaxis determined by the
static cavity birefringence, Zeeman splitting, and interaction-induced field,
while the acoustic contribution \(\mathbf B_a(t)\) is treated as an external
coherent drive. Including \(\mathbf B_a(t)\) inside the relaxation field would
produce additional terms of order \(\lambda\beta_h\), which are not essential
for the effects discussed in the main text.

Collecting the coherent precession, gain and loss, anisotropic decay, spin
relaxation, and reservoir dynamics, we obtain
\begin{align}
\partial_t \mathbf S
&=
(Rn_R-\gamma)\mathbf S
-\gamma_a |\mathbf S|\,\mathbf e_x
+
\left[
\mathbf B_0(\mathbf S)+\mathbf B_a(t)
\right]\times\mathbf S
-\lambda\,
\mathbf S\times
\left[
\mathbf B_0(\mathbf S)\times\mathbf S
\right],
\label{eq:supp_final_S}
\\
\partial_t n_R
&=
P-\gamma_R n_R-Rn_R|\mathbf S|.
\label{eq:supp_final_nR}
\end{align}
Here
\begin{equation}
\mathbf B_0(\mathbf S)
=
-\delta\,\mathbf e_x
+
\left(\alpha_s S_z-\Delta\right)\mathbf e_z,
\end{equation}
and
\begin{equation}
\mathbf B_a(t)
=
\beta_h(t)
\left(
\mathbf e_x\cos 2\phi+\mathbf e_y\sin 2\phi
\right)
\sin\Theta(t),
\qquad
\dot\Theta(t)=\omega(t).
\end{equation}
For constant acoustic amplitude and frequency this reduces to
\[
\mathbf B_a(t)
=
\beta_h
\left(
\mathbf e_x\cos 2\phi+\mathbf e_y\sin 2\phi
\right)
\sin\omega t .
\]
Equations \eqref{eq:supp_final_S} and \eqref{eq:supp_final_nR} are the
effective pseudospin--reservoir equations used in the main text.

\section*{Supplementary Note 2: Dimensionless units and conversion to physical units}

In this Supplementary Note we summarize the normalization used in the numerical
simulations and explain how the dimensionless quantities entering the equations
are converted to physical units. This point is important because the equations
are integrated in dimensionless form, whereas the figures in the main text are
shown in physical energy units, such as \(\mathrm{meV}\) and \(\upmu\mathrm{eV}\),
and in physical time units, such as \(\mathrm{ps}\) and \(\mathrm{ns}\).

We denote by \(\mathbf B_E\) an effective magnetic field measured in energy
units. The corresponding physical coherent pseudospin equation is
\begin{equation}
\frac{d\mathbf S_{\rm phys}}{dt_{\rm phys}}
=
\frac{1}{\hbar}
\mathbf B_E(\mathbf S_{\rm phys},t_{\rm phys})
\times
\mathbf S_{\rm phys}
+\ldots ,
\label{eq:supp_phys_spin_units}
\end{equation}
where the dots denote gain, loss and relaxation terms when they are included.
Equivalently, at the spinor level one may write
\begin{equation}
i\hbar\frac{\partial \Psi_{\rm phys}}{\partial t_{\rm phys}}
=
\frac12
\mathbf B_E\cdot\boldsymbol{\sigma}\,
\Psi_{\rm phys}.
\label{eq:supp_phys_spinor_units}
\end{equation}
Thus an effective splitting \(B_E\) expressed in energy units corresponds to
the angular precession frequency \(B_E/\hbar\).

We introduce an energy scale \(E_0\) and the associated time scale
\begin{equation}
t_0=\frac{\hbar}{E_0}.
\label{eq:supp_t0_def}
\end{equation}
The dimensionless time is defined by
\begin{equation}
t=\frac{t_{\rm phys}}{t_0}
=
\frac{E_0}{\hbar}t_{\rm phys}.
\label{eq:supp_dimensionless_time}
\end{equation}
All effective-field components entering the dimensionless pseudospin equation
are measured in units of \(E_0\). We use
\begin{equation}
\delta=\frac{\delta_E}{E_0},
\qquad
\Delta=\frac{\Delta_Z}{E_0},
\qquad
\beta_h=\frac{\beta_{h,E}}{E_0},
\qquad
\gamma_a=\frac{\gamma_{a,E}}{E_0}.
\label{eq:supp_field_scaling}
\end{equation}
Here \(\delta_E\), \(\Delta_Z\), and \(\beta_{h,E}\) are the physical energy
splittings corresponding to the static linear-polarization splitting, the
Zeeman splitting, and the acoustic-wave-induced strain splitting, respectively.
The quantity \(\gamma_{a,E}\) is the energy-linewidth representation of the
polarization-dependent lifetime anisotropy used in the figures of the main
text.

With these definitions, Eq.~\eqref{eq:supp_phys_spin_units} takes the
dimensionless form
\begin{equation}
\frac{d\mathbf S}{dt}
=
\mathbf B(\mathbf S,t)\times\mathbf S+\ldots ,
\label{eq:supp_dimless_spin_units}
\end{equation}
which is the form used throughout the analytical derivations and numerical
calculations. In this sense, setting \(\hbar=1\) is equivalent to using the
rescaled variables \eqref{eq:supp_t0_def}--\eqref{eq:supp_field_scaling}.

The acoustic phase in physical units is
\begin{equation}
\Theta(t_{\rm phys})
=
\int^{t_{\rm phys}}\Omega(t')\,dt',
\end{equation}
where \(\Omega\) is the physical angular frequency of the acoustic drive.
In dimensionless variables this becomes
\begin{equation}
\Theta(t)=\int^t\omega(t')\,dt',
\qquad
\omega=t_0\Omega=\frac{\hbar\Omega}{E_0}.
\label{eq:supp_omega_scaling}
\end{equation}
Therefore the quantity plotted on the horizontal frequency axes of the figures
is the corresponding acoustic drive energy
\begin{equation}
\hbar\Omega=E_0\omega.
\label{eq:supp_drive_energy}
\end{equation}
Thus, if the axis is labeled by \(\omega\) in \(\upmu\mathrm{eV}\) or
\(\mathrm{meV}\), it should be understood as the energy \(E_0\omega\)
associated with the physical acoustic angular frequency, rather than as the
dimensionless angular frequency itself. Similarly, all figure labels for the acoustic-drive amplitude and
the lifetime anisotropy use the energy-scaled quantities
\[
\beta_{h,E}=E_0\beta_h,\qquad
\gamma_{a,E}=E_0\gamma_a .
\]
Here \(\beta_h\) and \(\gamma_a\) are the dimensionless parameters
entering the equations of motion, while \(\beta_{h,E}\) and
\(\gamma_{a,E}\) are the corresponding dimensional energy splittings
or energy-linewidth parameters displayed in the figures. The corresponding cyclic frequency is
\begin{equation}
f=\frac{\Omega}{2\pi}=\frac{E_0\omega}{h}.
\label{eq:supp_cyclic_frequency}
\end{equation}

For all dimensional figures in the main text we use
\begin{equation}
E_0=0.020678~\mathrm{meV}
=
20.678~\upmu\mathrm{eV},
\qquad
t_0=\frac{\hbar}{E_0}=31.8315~\mathrm{ps}.
\label{eq:supp_E0_t0_values}
\end{equation}
The corresponding cyclic-frequency unit is
\begin{equation}
f_0=\frac{E_0}{h}
=
\frac{1}{2\pi t_0}
\simeq 5.00~\mathrm{GHz}.
\label{eq:supp_f0_value}
\end{equation}
For example, the dimensionless value \(\delta=0.8\) corresponds to
\begin{equation}
\delta_E=0.8E_0
=
0.016542~\mathrm{meV}
=
16.54~\upmu\mathrm{eV},
\qquad
\frac{\delta_E}{h}\simeq 4.00~\mathrm{GHz}.
\label{eq:supp_delta_conversion}
\end{equation}
The acoustic strain-induced splitting is converted in the same way:
\begin{equation}
\beta_{h,E}=E_0\beta_h.
\label{eq:supp_beta_conversion}
\end{equation}
Thus
\begin{equation}
\beta_h=0.04
\quad\Longleftrightarrow\quad
\beta_{h,E}=0.827~\upmu\mathrm{eV},
\end{equation}
whereas
\begin{equation}
\beta_h=0.15
\quad\Longleftrightarrow\quad
\beta_{h,E}=3.10~\upmu\mathrm{eV}.
\end{equation}
The smooth acoustic turn-on time used in the simulations,
\(\tau_{\rm on}=240\), corresponds to
\begin{equation}
\tau_{\rm on}^{\rm phys}=240\,t_0\simeq 7640~\mathrm{ps}.
\label{eq:supp_tau_conversion}
\end{equation}

We now specify the density normalization. Let \(n_*\) be the density unit used
to convert the dimensionless spinor and pseudospin variables to physical ones:
\begin{equation}
\Psi_{\rm phys}=\sqrt{n_*}\Psi,
\qquad
\mathbf S_{\rm phys}=n_*\mathbf S,
\qquad
n_{\rm phys}=n_* n.
\label{eq:supp_density_scaling}
\end{equation}
We use lower-case densities such as \(n\) and \(|\mathbf S|\) for
dimensionless quantities. Physical areal densities are denoted by capital
letters.

The interaction-induced part of the effective field is written in physical
units as
\begin{equation}
B_{z,E}^{\rm int}
=
\alpha_{\rm phys}S_{z,\rm phys},
\end{equation}
where \(\alpha_{\rm phys}\) has units of energy times area. In dimensionless
variables this term is
\begin{equation}
B_z^{\rm int}=\alpha_s S_z,
\end{equation}
where
\begin{equation}
\alpha_s
=
\frac{\alpha_{\rm phys}n_*}{E_0},
\qquad
\alpha_{\rm phys}
=
\frac{\alpha_s E_0}{n_*}.
\label{eq:supp_alpha_conversion}
\end{equation}
This is the relation used to compare simulations with different dimensionless
values of \(\alpha_s\) at fixed physical interaction strength.

For the conservative calculations shown in the main text we use
\begin{equation}
\alpha_s=0.35,
\qquad
\alpha_{\rm phys}=0.905~\upmu\mathrm{eV}\,\upmu\mathrm{m}^2.
\end{equation}
Therefore the density unit is
\begin{equation}
n_*^{(c)}
=
\frac{\alpha_s E_0}{\alpha_{\rm phys}}
=
\frac{0.35\times 20.678~\upmu\mathrm{eV}}
{0.905~\upmu\mathrm{eV}\,\upmu\mathrm{m}^2}
\simeq
8.0~\upmu\mathrm{m}^{-2}.
\label{eq:supp_nstar_conservative}
\end{equation}
Consequently, the dimensionless initial spin lengths
\begin{equation}
|\mathbf S_0|=0.5,\ 1.0,\ 1.5
\end{equation}
correspond to the physical condensate densities
\begin{equation}
N_0=4,\ 8,\ 12~\upmu\mathrm{m}^{-2}.
\label{eq:supp_conservative_densities}
\end{equation}
Equivalently,
\begin{equation}
\alpha_{\rm phys}
=
\frac{0.35\,E_0}{8.0~\upmu\mathrm{m}^{-2}}
=
9.05\times10^{-4}~\mathrm{meV}\,\upmu\mathrm{m}^2
=
0.905~\upmu\mathrm{eV}\,\upmu\mathrm{m}^2.
\label{eq:supp_alpha_conservative}
\end{equation}

For the pumped--dissipative calculations the dimensionless interaction strength
is \(\alpha_s=0.5\). In order to keep the same physical interaction coefficient,
\begin{equation}
\alpha_{\rm phys}=0.905~\upmu\mathrm{eV}\,\upmu\mathrm{m}^2,
\end{equation}
we use the density conversion scale
\begin{equation}
n_*^{(d)}
=
\frac{0.5\,E_0}{\alpha_{\rm phys}}
=
\frac{0.5\times20.678~\upmu\mathrm{eV}}
{0.905~\upmu\mathrm{eV}\,\upmu\mathrm{m}^2}
\simeq
11.43~\upmu\mathrm{m}^{-2}.
\label{eq:supp_nstar_dissipative}
\end{equation}
This scale is used only to interpret the dimensionless reservoir simulations
in physical density units. The condensate density itself is not fixed by hand
in the dissipative calculations; it is generated dynamically by the reservoir
equations.

The dissipative rates are normalized by the same time scale \(t_0\). If
\(N_R=n_* n_R\) is the physical reservoir density and \(P_{\rm phys}\) is the
physical pump rate per unit area, then the dimensionless parameters are
\begin{equation}
\gamma=\gamma_{\rm phys}t_0,
\qquad
\gamma_R=\gamma_{R,\rm phys}t_0,
\qquad
R=R_{\rm phys}n_*t_0,
\qquad
P=\frac{P_{\rm phys}t_0}{n_*}.
\label{eq:supp_rate_scaling}
\end{equation}
In the dissipative simulations discussed in the main text we use
\begin{equation}
\gamma=1,
\qquad
\gamma_R=4,
\qquad
R=1.
\label{eq:supp_dissipative_params}
\end{equation}
The corresponding dimensionless condensation threshold is
\begin{equation}
P_{\rm th}
=
\frac{\gamma\gamma_R}{R}
=
4.
\label{eq:supp_pth_units}
\end{equation}
For \(\gamma_a=0\) and in the absence of the acoustic drive, the nonzero
steady state satisfies
\begin{equation}
n_{R0}=\frac{\gamma}{R},
\qquad
n_0=\frac{P-P_{\rm th}}{\gamma}.
\label{eq:supp_steady_density_units}
\end{equation}
Since \(\gamma=1\) in our simulations, this reduces to
\begin{equation}
n_0=P-P_{\rm th}.
\end{equation}
If the pump is labeled by the ratio
\begin{equation}
p=\frac{P}{P_{\rm th}},
\end{equation}
then the corresponding physical steady-state density estimate is
\begin{equation}
N_{\rm ss}(p)
=
n_*^{(d)}
\frac{P_{\rm th}(p-1)}{\gamma}.
\label{eq:supp_Nss_p}
\end{equation}
For the parameters used in the main text, this means
\begin{equation}
N_{\rm ss}(p)
\simeq
11.43~\upmu\mathrm{m}^{-2}\times 4(p-1)
=
45.72(p-1)~\upmu\mathrm{m}^{-2}.
\end{equation}

The analytic density-shift estimate used as a reference for the pump-dependent
resonance shift follows from the linearized spin dynamics around the
linearly polarized state. For \(\Delta=0\), without the acoustic drive,
\begin{equation}
\mathbf B_0=(-\delta,0,\alpha_s S_z).
\end{equation}
Linearizing around
\begin{equation}
\mathbf S=(n,S_y,S_z),
\qquad
|S_y|,|S_z|\ll n,
\end{equation}
one obtains
\begin{equation}
\dot S_y=(\delta+\alpha_s n)S_z,
\qquad
\dot S_z=-\delta S_y.
\label{eq:supp_linearized_spin_units}
\end{equation}
Therefore the small-oscillation frequency is
\begin{equation}
\omega_{\rm lin}^2
=
\delta(\delta+\alpha_s n).
\label{eq:supp_omega_lin_units}
\end{equation}
Restoring physical units gives the characteristic energy
\begin{equation}
E_{\rm an}
=
E_0\omega_{\rm lin}
=
\sqrt{
\delta_E
\left[
\delta_E+\alpha_{\rm phys}N
\right]
}.
\label{eq:supp_Ean_units}
\end{equation}
For the pump scan we use \(N=N_{\rm ss}(p)\), with \(N_{\rm ss}(p)\) given by
Eq.~\eqref{eq:supp_Nss_p}. This curve is only a small-oscillation estimate and
is not a fit to the numerical hysteresis branches, which are determined by
finite-amplitude driven dynamics, reservoir feedback, and the history of the
sweep.

Finally, the response amplitude plotted in the main text is dimensionless:
\begin{equation}
A_j(W)
=
\frac12
\left[
\max_{t\in W}\frac{S_j(t)}{|\mathbf S(t)|}
-
\min_{t\in W}\frac{S_j(t)}{|\mathbf S(t)|}
\right],
\qquad j=x,y,z.
\label{eq:supp_Aj_units}
\end{equation}
The window \(W\) is either a late-time window of an independent
frequency-response calculation or a sliding window along a slow sweep. Physical
time intervals are obtained by multiplying the corresponding dimensionless
intervals by \(t_0\).

\section*{Supplementary Note 3: Conservative dynamics in terms of the circular-component amplitudes}

In this Supplementary Note we rewrite the conservative pseudospin dynamics in
terms of the circularly polarized condensate amplitudes \(\psi_\pm\). This
representation is useful for comparison with the standard two-component
Gross--Pitaevskii form and makes explicit how the static linear-polarization
splitting and the acoustic strain-induced splitting couple the two circular
components.

We consider the conservative limit of the model derived in Supplementary
Note~1. Thus the coupling to the reservoir, radiative decay,
polarization-dependent loss, and spin relaxation are neglected:
\begin{equation}
R=\gamma=\gamma_a=\lambda=0.
\end{equation}
The condensate dynamics is then governed by the Hermitian spinor equation
\begin{equation}
i\partial_t\Psi=\hat H_{\mathrm{eff}}\Psi,
\qquad
\Psi=
\begin{pmatrix}
\psi_+\\
\psi_-
\end{pmatrix},
\label{eq:supp3_spinor_cons}
\end{equation}
where all quantities are written in the dimensionless units introduced in
Supplementary Note~2. The effective Hamiltonian is
\begin{equation}
\hat H_{\mathrm{eff}}
=
\frac12
\left[
\mathbf B_0(\mathbf S)+\mathbf B_a(t)
\right]\cdot\boldsymbol{\sigma}.
\label{eq:supp3_Heff_cons}
\end{equation}
Here
\begin{equation}
\mathbf B_0(\mathbf S)
=
-\delta\,\mathbf e_x
+
\left(\alpha_s S_z-\Delta\right)\mathbf e_z,
\label{eq:supp3_B0_cons}
\end{equation}
and
\begin{equation}
\mathbf B_a(t)
=
\beta_h(t)
\left(
\mathbf e_x\cos 2\phi+\mathbf e_y\sin 2\phi
\right)
\sin\Theta(t),
\qquad
\dot\Theta(t)=\omega(t).
\label{eq:supp3_Ba_general}
\end{equation}
For fixed acoustic amplitude and frequency this reduces to
\begin{equation}
\mathbf B_a(t)
=
\beta_h
\left(
\mathbf e_x\cos 2\phi+\mathbf e_y\sin 2\phi
\right)
\sin\omega t.
\label{eq:supp3_Ba_fixed}
\end{equation}
The \(z\)-component of the pseudospin is
\begin{equation}
S_z=|\psi_+|^2-|\psi_-|^2.
\label{eq:supp3_Sz}
\end{equation}

Introducing the Cartesian components of the total effective field,
\begin{equation}
\mathbf B(t,\mathbf S)
=
\mathbf B_0(\mathbf S)+\mathbf B_a(t),
\end{equation}
we have, for the fixed-frequency case,
\begin{align}
B_x
&=
-\delta+\beta_h\cos 2\phi\,\sin\omega t,
\\
B_y
&=
\beta_h\sin 2\phi\,\sin\omega t,
\\
B_z
&=
\alpha_s\left(|\psi_+|^2-|\psi_-|^2\right)-\Delta.
\end{align}
The Hamiltonian \eqref{eq:supp3_Heff_cons} therefore takes the matrix form
\begin{equation}
\hat H_{\mathrm{eff}}
=
\frac12
\begin{pmatrix}
B_z & B_x-iB_y\\
B_x+iB_y & -B_z
\end{pmatrix}.
\label{eq:supp3_Heff_matrix}
\end{equation}
Consequently, the circular components obey
\begin{align}
i\partial_t \psi_+
&=
\frac12
\left[
\alpha_s
\left(|\psi_+|^2-|\psi_-|^2\right)
-\Delta
\right]\psi_+
+
\frac12
\left[
-\delta+\beta_h e^{-2i\phi}\sin\omega t
\right]\psi_-,
\label{eq:supp3_psi_plus_general}
\\
i\partial_t \psi_-
&=
-\frac12
\left[
\alpha_s
\left(|\psi_+|^2-|\psi_-|^2\right)
-\Delta
\right]\psi_-
+
\frac12
\left[
-\delta+\beta_h e^{2i\phi}\sin\omega t
\right]\psi_+.
\label{eq:supp3_psi_minus_general}
\end{align}
Equations \eqref{eq:supp3_psi_plus_general} and
\eqref{eq:supp3_psi_minus_general} show explicitly that the static
linear-polarization splitting \(\delta\) produces a time-independent coherent
coupling between the two circular components, while the acoustic wave produces
a time-periodic modulation of this coupling. The phase of the acoustic coupling
is determined by twice the acoustic propagation angle, \(2\phi\). The nonlinear
term proportional to \(\alpha_s\) produces a self-induced circular-polarization
splitting governed by the population imbalance \(S_z\).

Since the Hamiltonian \eqref{eq:supp3_Heff_matrix} is Hermitian, the total
condensate density
\begin{equation}
n=|\psi_+|^2+|\psi_-|^2
\end{equation}
is conserved:
\begin{equation}
\partial_t n=0.
\label{eq:supp3_density_cons}
\end{equation}
Thus, in the conservative regime the dynamics reduces to nonlinear coherent
oscillations of the spinor state at fixed pseudospin length,
\(|\mathbf S|=n=\mathrm{const}\).

The main text focuses on the geometry
\begin{equation}
\phi=\frac{\pi}{4},
\end{equation}
for which the acoustic field is perpendicular to the static
linear-polarization splitting:
\begin{equation}
\mathbf B_a(t)=\beta_h\,\mathbf e_y\sin\omega t.
\label{eq:supp3_Ba_phi45}
\end{equation}
In this case
\[
e^{-2i\phi}=e^{-i\pi/2}=-i,
\qquad
e^{2i\phi}=e^{i\pi/2}=i,
\]
and Eqs.~\eqref{eq:supp3_psi_plus_general}--\eqref{eq:supp3_psi_minus_general}
become
\begin{align}
i\partial_t \psi_+
&=
\frac12
\left[
\alpha_s
\left(|\psi_+|^2-|\psi_-|^2\right)
-\Delta
\right]\psi_+
+
\frac12
\left(
-\delta-i\beta_h\sin\omega t
\right)\psi_-,
\label{eq:supp3_psi_plus_phi45}
\\
i\partial_t \psi_-
&=
-\frac12
\left[
\alpha_s
\left(|\psi_+|^2-|\psi_-|^2\right)
-\Delta
\right]\psi_-
+
\frac12
\left(
-\delta+i\beta_h\sin\omega t
\right)\psi_+.
\label{eq:supp3_psi_minus_phi45}
\end{align}
Finally, in the case without external magnetic field, \(\Delta=0\), this gives
\begin{align}
i\partial_t \psi_+
&=
\frac{\alpha_s}{2}
\left(|\psi_+|^2-|\psi_-|^2\right)\psi_+
+
\frac12
\left(
-\delta-i\beta_h\sin\omega t
\right)\psi_-,
\label{eq:supp3_psi_plus_phi45_noDelta}
\\
i\partial_t \psi_-
&=
-\frac{\alpha_s}{2}
\left(|\psi_+|^2-|\psi_-|^2\right)\psi_-
+
\frac12
\left(
-\delta+i\beta_h\sin\omega t
\right)\psi_+.
\label{eq:supp3_psi_minus_phi45_noDelta}
\end{align}
This form makes clear that, in the \(\phi=\pi/4\) geometry, the acoustic wave
acts as a transverse coherent drive of the pseudospin, whereas the
polariton--polariton interaction produces a density-dependent nonlinear
detuning through the self-induced circular splitting.

\section*{Supplementary Note 4: Linearized spin dynamics under acoustic driving}

In this Supplementary Note we analyze the small-amplitude pseudospin response
to the acoustic drive in the simplest pumped--dissipative regime. The purpose is
to derive the acoustic spin-resonance condition and to separate the roles of the
reservoir-induced gain saturation and the Gilbert-type spin relaxation.

We consider the case
\begin{equation}
\Delta=0,
\qquad
\gamma_a=0,
\end{equation}
and choose the acoustic-wave propagation direction as
\begin{equation}
\phi=\frac{\pi}{4}.
\end{equation}
Then the acoustic field is directed along the \(y\) axis:
\begin{equation}
\mathbf B_a(t)=\beta_h\,\mathbf e_y\sin\omega t.
\label{eq:supp4_Ba_y}
\end{equation}
The stationary effective field is
\begin{equation}
\mathbf B_0(\mathbf S)
=
-\delta\,\mathbf e_x+\alpha_s S_z\,\mathbf e_z.
\label{eq:supp4_B0_caseA}
\end{equation}
The dynamical equations reduce to
\begin{align}
\partial_t \mathbf S
&=
(Rn_R-\gamma)\mathbf S
+
\left[
\mathbf B_0(\mathbf S)+\mathbf B_a(t)
\right]\times\mathbf S
-\lambda\,
\mathbf S\times
\left[
\mathbf B_0(\mathbf S)\times\mathbf S
\right],
\label{eq:supp4_S_caseA}
\\
\partial_t n_R
&=
P-\gamma_R n_R-Rn_R|\mathbf S|.
\label{eq:supp4_nR_caseA}
\end{align}

In the absence of the acoustic drive, \(\beta_h=0\), we look for a stationary
state linearly polarized along the \(X\) axis:
\begin{equation}
\mathbf S_0=n_0\mathbf e_x,
\qquad
n_{R0}=\mathrm{const}.
\label{eq:supp4_stationary_state}
\end{equation}
For this state \(S_z=0\), and hence
\begin{equation}
\mathbf B_0(\mathbf S_0)=-\delta\,\mathbf e_x.
\end{equation}
Thus \(\mathbf S_0\) is collinear with \(\mathbf B_0(\mathbf S_0)\), so both the
precession term and the Gilbert-type relaxation term vanish. The condensate
equation then gives the gain--loss balance condition
\begin{equation}
(Rn_{R0}-\gamma)\mathbf S_0=0.
\end{equation}
For a nonzero condensate this yields
\begin{equation}
n_{R0}=\frac{\gamma}{R}.
\label{eq:supp4_nR0}
\end{equation}
Substituting Eq.~\eqref{eq:supp4_nR0} into the reservoir equation gives
\begin{equation}
0
=
P-\gamma_R n_{R0}-Rn_{R0}n_0,
\end{equation}
and therefore
\begin{equation}
n_0
=
\frac{P}{\gamma}
-
\frac{\gamma_R}{R}
=
\frac{P-P_{\rm th}}{\gamma},
\label{eq:supp4_n0}
\end{equation}
where
\begin{equation}
P_{\rm th}
=
\frac{\gamma\gamma_R}{R}
\label{eq:supp4_Pth}
\end{equation}
is the condensation threshold. Thus the reservoir fixes the threshold and the
stationary density. The spin relaxation term does not determine the density;
instead, it damps transverse pseudospin motion and allows the driven system to
approach stable long-time states.

To study the weak acoustic response, we linearize around the stationary state
using perturbations \(\xi_i\) for the pseudospin and \(\nu\) for the reservoir:
\begin{equation}
\mathbf S
=
(n_0+\xi_x)\mathbf e_x+\xi_y\mathbf e_y+\xi_z\mathbf e_z,
\qquad
n_R=n_{R0}+\nu,
\label{eq:supp4_linearization_ansatz}
\end{equation}
where
\begin{equation}
|\xi_x|,\ |\xi_y|,\ |\xi_z|,\ |\nu|\ll n_0.
\end{equation}
This notation keeps the small perturbations distinct from the normalized
pseudospin notation introduced locally in Supplementary Note~5. To linear order,
\begin{equation}
|\mathbf S|\simeq n_0+\xi_x,
\end{equation}
and
\begin{equation}
\mathbf B_0(\mathbf S)
\simeq
-\delta\,\mathbf e_x+\alpha_s \xi_z\,\mathbf e_z.
\end{equation}

Keeping terms linear in \(\xi_x,\xi_y,\xi_z,\nu\) and in the acoustic-drive
amplitude \(\beta_h\), we find that the longitudinal density--reservoir sector
decouples from the transverse spin sector:
\begin{align}
\partial_t \xi_x
&=
n_0R\,\nu,
\label{eq:supp4_sx}
\\
\partial_t \nu
&=
-(\gamma_R+Rn_0)\nu-\gamma \xi_x.
\label{eq:supp4_nu}
\end{align}
The transverse pseudospin dynamics is governed by
\begin{align}
\partial_t \xi_y
&=
(\delta+\alpha_s n_0)\xi_z
-
\lambda n_0\delta\,\xi_y,
\label{eq:supp4_sy}
\\
\partial_t \xi_z
&=
-\delta \xi_y
-
\lambda n_0(\delta+\alpha_s n_0)\xi_z
-
\beta_h n_0\sin\omega t.
\label{eq:supp4_sz}
\end{align}
The acoustic drive enters Eq.~\eqref{eq:supp4_sz} because
\[
\beta_h\mathbf e_y\times n_0\mathbf e_x
=
-\beta_h n_0\,\mathbf e_z.
\]

Equations \eqref{eq:supp4_sx} and \eqref{eq:supp4_nu} show that, in this
minimal case, the reservoir controls the condensate density but does not enter
the linear transverse spin resonance directly. The resonance frequency is fixed
by the dimensionless density \(n_0\), which is itself set by the pump through
Eq.~\eqref{eq:supp4_n0}.

Eliminating \(\xi_y\) from Eqs.~\eqref{eq:supp4_sy} and
\eqref{eq:supp4_sz}, we obtain a driven damped oscillator equation for
\(\xi_z\):
\begin{align}
\partial_t^2 \xi_z
&+
\lambda n_0(2\delta+\alpha_s n_0)\partial_t \xi_z
+
(1+\lambda^2 n_0^2)
\delta(\delta+\alpha_s n_0)\xi_z
\nonumber\\
&=
-\beta_h n_0
\left[
\omega\cos\omega t
+
\lambda n_0\delta\sin\omega t
\right].
\label{eq:supp4_sz_oscillator}
\end{align}
Thus the acoustic wave drives a damped pseudospin oscillator.

In the weak-relaxation limit, \(\lambda n_0\ll1\), the dimensionless
small-oscillation frequency is
\begin{equation}
\omega_s
\simeq
\sqrt{
\delta(\delta+\alpha_s n_0)
}.
\label{eq:supp4_omega_s}
\end{equation}
The corresponding damping rate is
\begin{equation}
\Gamma_s
\simeq
\frac{\lambda n_0}{2}
(2\delta+\alpha_s n_0).
\label{eq:supp4_Gamma_s}
\end{equation}
The acoustic spin-resonance condition is therefore
\begin{equation}
\omega\simeq\omega_s.
\label{eq:supp4_resonance}
\end{equation}
Restoring physical units according to Supplementary Note~2 gives the resonance
energy estimate
\begin{equation}
E_s
=
E_0\omega_s
=
\sqrt{
\delta_E
\left[
\delta_E+\alpha_{\rm phys}N_{\rm st}
\right]
},
\label{eq:supp4_Es_phys}
\end{equation}
where \(N_{\rm st}=n_*^{(d)}n_0\) is the physical steady-state condensate
density corresponding to the dimensionless density \(n_0\). This notation
avoids using \(N_0\) for the pumped--dissipative stationary density; \(N_0\) is
reserved for the physical initial density in the conservative simulations.

The \(X\)-polarized stationary state is linearly stable with respect to
transverse spin perturbations provided
\begin{equation}
\delta(\delta+\alpha_s n_0)>0
\label{eq:supp4_stability}
\end{equation}
and the damping coefficient in Eq.~\eqref{eq:supp4_sz_oscillator} is positive,
\begin{equation}
\lambda n_0(2\delta+\alpha_s n_0)>0.
\end{equation}
For the parameter regime used in the main text, \(\delta>0\),
\(\alpha_s>0\), \(n_0>0\), and \(\lambda>0\), both conditions are satisfied.

\newpage
\section*{Supplementary Note 5: Stationary states with lifetime anisotropy}

In this Supplementary Note we summarize the stationary equations used to
interpret the lifetime-anisotropy-induced bifurcation discussed in the main
text. We consider the system without acoustic drive,
\(\mathbf B_a=0\), and set \(\Delta=0\). The stationary effective magnetic field
is then
\begin{equation}
\mathbf B_0(\mathbf S)=(-\delta,0,\alpha_s S_z).
\label{eq:supp5_B0}
\end{equation}
It is convenient to write the pseudospin as
\begin{equation}
\mathbf S=n\mathbf s,
\qquad
\mathbf s=(s_x,s_y,s_z),
\qquad
|\mathbf s|=1,
\label{eq:supp5_normalized_spin}
\end{equation}
where \(n=|\mathbf S|\) is the condensate density. The stationary reservoir
balance gives
\begin{equation}
n_R=\frac{P}{\gamma_R+Rn}.
\label{eq:supp5_nR_stat}
\end{equation}
Taking the scalar product of Eq.~\eqref{eq:supp_final_S} with
\(\mathbf S\), and using the fact that the precession and Gilbert terms do not
change \(|\mathbf S|\), one obtains the stationary density condition
\begin{equation}
Rn_R-\gamma=\gamma_a s_x.
\label{eq:supp5_density_balance}
\end{equation}
Equations \eqref{eq:supp5_nR_stat} and \eqref{eq:supp5_density_balance}
therefore determine the condensate density for a given orientation \(s_x\).
For \(\gamma_a=0\) they reduce to the usual isotropic-threshold result,
\(n=(P-P_{\rm th})/\gamma\), with
\(P_{\rm th}=\gamma\gamma_R/R\).

The remaining stationary conditions determine the pseudospin orientation. Using
Eq.~\eqref{eq:supp5_density_balance}, the stationary pseudospin equation can be
written as
\begin{equation}
0=
\gamma_a(s_x\mathbf s-\mathbf e_x)
+
\mathbf B_0(n\mathbf s)\times\mathbf s
-
\lambda n\,
\mathbf s\times
\left[\mathbf B_0(n\mathbf s)\times\mathbf s\right].
\label{eq:supp5_orientation_vector}
\end{equation}
This form makes explicit that the lifetime anisotropy competes with coherent
precession and spin relaxation at fixed pseudospin length. For
\(\Delta=0\), let
\begin{equation}
A=\alpha_s n,
\qquad
b=\mathbf s\cdot\mathbf B_0=-\delta s_x+A s_z^2 .
\label{eq:supp5_A_b}
\end{equation}
Then Eq.~\eqref{eq:supp5_orientation_vector} is equivalent to the algebraic
system
\begin{align}
0&=
\gamma_a(s_x^2-1)-A s_zs_y
-
\lambda n(-\delta-b s_x),
\label{eq:supp5_stat_x}
\\
0&=
\gamma_a s_xs_y+s_z(As_x+\delta)+\lambda n b s_y,
\label{eq:supp5_stat_y}
\\
0&=
\gamma_a s_xs_z-\delta s_y
-
\lambda n s_z(A-b),
\label{eq:supp5_stat_z}
\\
1&=s_x^2+s_y^2+s_z^2.
\label{eq:supp5_norm}
\end{align}
Together with Eqs.~\eqref{eq:supp5_nR_stat} and
\eqref{eq:supp5_density_balance}, these equations form the stationary problem
solved numerically to obtain the maps of
\(|\langle S_z/|\mathbf S|\rangle|\) shown in the main text.

At \(\Delta=0\), the stationary equations are invariant under
\begin{equation}
(s_x,s_y,s_z)\rightarrow(s_x,-s_y,-s_z).
\label{eq:supp5_symmetry}
\end{equation}
Therefore any finite-\(s_z\) solution appears together with a symmetry-related
partner of opposite circular polarization. The color maps in the main text show
\(|\langle S_z/|\mathbf S|\rangle|\) because the two branches are equivalent in
the absence of Zeeman splitting. The role of \(\gamma_a\) is to create the
polarization-dependent dissipative imbalance, while the nonlinear field
\(\alpha_s S_z\mathbf e_z\) and the spin-relaxation term determine whether the
finite-\(S_z\) branches are stable for a given pump and lifetime anisotropy.
This explains why increasing \(\lambda\) can suppress the out-of-plane
stationary component in the parameter region used in the main text.

For the bifurcated-state and switching simulations in the main text, the
branch was selected by a fixed undriven-relaxation protocol. The system was
first propagated at \(\beta_h=0\) from a branch-selecting initial condition with
normalized circular-polarization component
\(S_z/|\mathbf S|\simeq0.24\), as displayed in the inset of the
lifetime-anisotropy figure in the main text. This undriven relaxation selects
the stationary branch with \(S_z/|\mathbf S|\simeq-0.543\). The
symmetry-related branch is obtained by reversing the sign of the initial
circular-polarization seed. Because the maps in the main text show
\(|\langle S_z/|\mathbf S|\rangle|\), the sign choice does not affect the
stationary maps, but it fixes the initial branch for the switching traces.

\section*{Supplementary Note 6: Small-oscillation guide for Zeeman tuning}

In this Supplementary Note we derive the small-oscillation guide used in the
Zeeman-tuning map. The numerical map in the main text is calculated from the
full pumped--dissipative equations, but the guide is obtained by linearizing the
conservative precession dynamics in the stationary effective field around the
undriven pumped--dissipative stationary state \(\mathbf S_{\rm st}\). Thus we
consider
\begin{equation}
\dot{\mathbf S}=\mathbf B_0(\mathbf S)\times\mathbf S,
\qquad
\mathbf B_0(\mathbf S)=(-\delta,0,\alpha_s S_z-\Delta).
\label{eq:supp6_cons_spin}
\end{equation}
The component form is
\begin{align}
\dot S_x&=-(\alpha_s S_z-\Delta)S_y,
\label{eq:supp6_full_x}
\\
\dot S_y&=(\alpha_s S_z-\Delta)S_x+\delta S_z,
\label{eq:supp6_full_y}
\\
\dot S_z&=-\delta S_y.
\label{eq:supp6_full_z}
\end{align}
We write the small perturbation as \(\boldsymbol\eta\), rather than
\(\mathbf s\), to avoid confusion with the normalized pseudospin notation used
locally in Supplementary Note~5:
\begin{equation}
\mathbf S=\mathbf S_{\rm st}+\boldsymbol\eta,
\qquad
|\boldsymbol\eta|\ll |\mathbf S_{\rm st}|.
\end{equation}
We then define
\begin{equation}
B_z^{\rm st}=\alpha_s S_z^{\rm st}-\Delta.
\label{eq:supp6_Bzst}
\end{equation}
Keeping terms linear in \(\boldsymbol\eta\) gives
\begin{equation}
\frac{d}{dt}
\begin{pmatrix}
\eta_x\\ \eta_y\\ \eta_z
\end{pmatrix}
=J
\begin{pmatrix}
\eta_x\\ \eta_y\\ \eta_z
\end{pmatrix},
\label{eq:supp6_linear_system}
\end{equation}
with
\begin{equation}
J=
\begin{pmatrix}
0 & -B_z^{\rm st} & -\alpha_s S_y^{\rm st}\\
B_z^{\rm st} & 0 & \delta+\alpha_s S_x^{\rm st}\\
0 & -\delta & 0
\end{pmatrix}.
\label{eq:supp6_J}
\end{equation}
For the Zeeman-tuning simulations in the main text, \(\gamma_a=0\) and the
undriven stationary state lies in the \((S_x,S_z)\) plane, so that
\(S_y^{\rm st}=0\). In this case the matrix has one zero eigenvalue and a pair
of imaginary eigenvalues,
\begin{equation}
\lambda_\pm=\pm i\omega_{\rm guide},
\label{eq:supp6_lambdas}
\end{equation}
where
\begin{equation}
\omega_{\rm guide}^2
=
\left(B_z^{\rm st}\right)^2
+
\delta\left(\delta+\alpha_s S_x^{\rm st}\right).
\label{eq:supp6_guide_frequency}
\end{equation}
The guide plotted in the main text is displayed in energy units as
\begin{equation}
E_{\rm guide}=E_0\omega_{\rm guide}.
\label{eq:supp6_Eguide}
\end{equation}
This estimate captures the local small-oscillation frequency around the
undriven stationary state. It is not expected to coincide exactly with the
numerically extracted resonance ridge, because the latter is obtained from a
finite-amplitude acoustically driven trajectory in the full nonlinear
pumped--dissipative system.

\bibliography{main}